\documentclass[prd,
,superscriptaddress,
nofootinbib,%
tightenlines ]{revtex4}
\usepackage{epsfig}
\usepackage[colorlinks=true,linkcolor=blue,urlcolor=blue,citecolor=blue]{hyperref}
\usepackage{amsfonts}
\usepackage{amssymb}
\usepackage{amsmath}
\usepackage{slashed}

\newcommand{\ben}{\begin{displaymath}}
\newcommand{\een}{\end{displaymath}}
\newcommand{\be}{\begin{equation}}
\newcommand{\ee}{\end{equation}}
\newcommand{\bea}{\begin{eqnarray}}
\newcommand{\eea}{\end{eqnarray}}

\usepackage[toc,page]{appendix}
\usepackage[ngerman, english]{babel}
\usepackage{pstricks}
\usepackage{color}
\usepackage{caption}

\usepackage{float}
\usepackage{mwe}
\usepackage{subfigure}

\usepackage{pdfpages}
\usepackage{hyperref}
\graphicspath{ {./images/} }

\begin{document}
\title{Gravitational form factors of the nucleon and one pion graviproduction in chiral EFT}

\title{Gravitational form factors of the nucleon and one pion graviproduction in chiral EFT}

\author{H.~Alharazin}
\affiliation{Institut f\"ur Theoretische Physik II, Ruhr-Universit\"at Bochum,  D-44780 Bochum,
 Germany}

\begin{abstract}
In the framework of chiral effective field theory of delta resonances, nucleons and pions interacting with background gravitational field we calculate the gravitational form factors of the nucleon up to fourth order in the small scale expansion and obtain the  long-range behavior of the corresponding contributions to the energy, spin, pressure and shear force distributions.~By comparing nucleon gravitational form factors with and without delta contributions we conclude that explicit inclusion of deltas plays an important role.~Next we explore the Lorentz structure of the $N \mapsto N \pi$  transition matrix element of the conserved symmetric energy-momentum tensor and introduce its parametrization in terms of twelve transition form factors.~We use the chiral effective field theory to calculate the tree-order contributions to the gravitational transition form factors of the pion graviproduction off the nucleon up to third order.
\end{abstract}

\maketitle

\section{Introduction}
Hadronic matrix elements of the energy-momentum tensor (EMT) have been studied intensively over the past decades on both theoretical and experimental sides.~These quantities, which can be extracted from the GPDs \cite{Muller:1994ses,Ji:1996ek,Radyushkin:1996nd,Goeke:2001tz}, encode global characteristics of hadrons, such as the charge, mass, spin and the D-term \cite{Kobzarev:1962wt,Pagels:1966zza,Polyakov:2002yz,Polyakov:2018zvc}.~The gravitational form factors (GFFs) of the nucleon, which are accessed experimentally in exclusive processes like deeply virtual Compton scattering (DVCS) \cite{Ji:1996ek,Radyushkin:1997ki} and hard exclusive meson production \cite{Collins:1996fb}, have been the main focus of many works in the last years, see, e.g., recent colloquium \cite{Burkert:2023wzr} and references therein.
~Moreover, non-symmetric hadronic matrix elements of EMT, which are parametrized in terms of gravitational transition form factors (GTFFs), and their connection to the transition GPDs have awaken increasing interest in the past few years and started to be studied both theoretically and experimentally, 
see, e.g., Refs.~\cite{Kim:2022bwn,Semenov-Tian-Shansky:2023bsy,Kim:2023xvw,Kroll:2022roq}.

    Experimentally, the GFFs and the GTFFs are hard to obtain at low energies, so we may rely on chiral effective field theory (EFT) to make predictions for these quantities.~In Ref.~\cite{Alharazin:2020yjv}, the effective chiral Lagrangian of nucleons and pions has been extended to curved spacetime up to second chiral order and it was used to obtain the GFFs of the nucleon to fourth order of chiral counting.~In the first part of the current work we extend the Lagrangian of Ref.~\cite{Alharazin:2020yjv} by including the terms with delta resonances at the leading order and the order three terms of coupling to external gravitational field relevant for calculations of the current work.~Next we calculate the nucleon GFFs up to fourth order in the small scale expansion \cite{Hemmert:1996xg,Hemmert:1997ye} and obtain the long-range behavior of the corresponding contributions to the spatial densities, i.e. to the energy, spin, pressure and shear force distributions.~In the second part we discuss the one pion graviproduction (OPGP) off the nucleon, in which the initial nucleon scatters on external gravitational field and emits a pion in the final state.~The OPGP, which can be accessed in hard exclusive processes \cite{Max:1,Guichon:2003ah,Max:2,Kivel:2004bb}, was discussed for the first time in Ref.~\cite{Max:1} where the authors showed that it is related to the second moment of the $N \pi$-distributions.~They also showed that soft-pion theorems, which are related to OPGP at low energies, can be used to provide useful estimates of the soft-pion contamination in hard exclusive reactions and they can serve as an additional source of information about GPDs.~At the threshold, the matrix element of OPGP is parametrized in terms of three independent GTFFs, that were introduced in Ref.~\cite{Polyakov:2020rzq}.~The OPGP in the framework of chiral EFT at the threshold was also studied in Ref.~\cite{Alharazin:2020yjv}.~We extend the work of Refs.~\cite{Alharazin:2020yjv,Polyakov:2020rzq} beyond the threshold and give a general parametrization for the matrix element of a conserved EMT corresponding to OPGP in terms of twelve independent GTFFs.~Next, using the chiral effective Lagrangian, we calculate tree-level diagrams contributing to the GTFFs up to third order.

        This work is organized as follows: In section~\ref{I} we specify the relevant terms of the  effective Lagrangian for pions, nucleons and delta resonances in curved spacetime and the corresponding expression for the EMT.~In section~\ref{II} we discuss renormalization of the loop diagrams contributing in our calculations and obtain the nucleon GFFs at fourth order in the small scale expansion.~Next, in section \ref{Sec:2} we discuss  the parametrization of the matrix element of OPGP and use the Lagrangians from section \ref{I} to obtain the tree-order contributions to the GTFFs.~The results of the work are summarized in section \ref{III}.~In appendix \ref{A-I} we show the building blocks of the action together with the corresponding expressions of the EMT and appendix \ref{A-II} contains variations of terms in the action with gravity.~In appendices \ref{ros} and \ref{ros-I} are given identities, which we used to reduce some redundant structures in the parametrization of OPGP and also the tree-order expressions of the GTFFs.
        
\clearpage
\section{Effective action in curved spacetime and the energy-momentum tensor}
\label{I}
The action corresponding to the effective Lagrangian of pions, nucleons and delta resonances
in the presence of external gravitational field is obtained from the corresponding expressions in the flat spacetime.~The terms relevant for our calculation have the following form:
\begin{eqnarray}
S_{\rm \pi}^{(2)} 
&=& \int d^4x \sqrt{-g}\, \left\{ \frac {F_\pi^2}{4}\,  {\rm Tr}
( D_\mu U  (D^\mu U)^\dagger ) + \frac{F_\pi^2}{4}\,{\rm Tr}(\chi U^\dagger +U \chi^\dagger) \right\}\,,
\label{PionAction} 
\\
S_{\rm N \pi}^{(1,2,3)}  
& = & 
\int d^4x \sqrt{-g}\, \biggl\{\, \bar\Psi \, i  \gamma^\mu
\overset{\leftrightarrow}{\nabla}_\mu \Psi -m \bar\Psi\Psi  +\frac{g_A}{2}\, \bar\Psi \gamma^\mu \gamma_5 u_\mu \Psi  +  c_1 \langle \chi_+\rangle  \bar\Psi  \Psi\nonumber
\\
&-& \frac{c_2}{8 m^2} g^{\mu\alpha} g^{\nu\beta} \langle u_\mu u_\nu\rangle  \left( \bar\Psi \left\{ \nabla_\alpha, \nabla_\beta\right\}  \Psi+
 \left\{ \nabla_\alpha, \nabla_\beta\right\}  \bar\Psi \Psi \right) + \frac{c_3}{2} \, g^{\mu\nu} \langle u_\mu u_\nu\rangle  \bar\Psi  \Psi  \nonumber\\
&+&  \frac{i c_4}{4} \, \bar\Psi \,e^\mu_a e^\nu_b \sigma^{ab}\,\left[ u_\mu ,u_\nu\right] \Psi 
+ c_5 \bar\Psi\hat \chi_+ \Psi + \frac{c_6}{8m} \bar\Psi \,e^\mu_a e^\nu_b \sigma^{ab} F^+_{\mu\nu} \Psi + \frac{c_7}{8m} \bar \Psi \,e^\mu_a e^\nu_b \sigma^{ab} 
\langle F^+_{\mu\nu} \rangle \Psi
\nonumber
\\
&+&\frac{1}{2} \tilde d_{10} \bar \Psi  \gamma^\mu \gamma_5  \langle u^2 \rangle u_\mu \Psi 
+ \frac{1}{2}  \tilde d_{16} \bar \Psi  \gamma^\mu \gamma_5  \langle \chi_+ \rangle u_\mu \Psi 
+ \frac{1}{2} \tilde d_{17} \bar \Psi  \gamma^\mu \gamma_5  \langle \chi_+ u_\mu \rangle  \Psi+\frac{i}{2} \tilde d_{18} \bar \Psi  \gamma^\mu \gamma_5  \left[ D_\mu ,  \chi_- \right]  \Psi \biggr\} \,, 
\label{PiNAction} 
\\ 
S_{\Delta \pi}^{(1)} 
& = &
 - \int d^4 x  \sqrt{-g} \biggl\{  
\Bar{\Psi}^{i \mu}  \,  i \gamma^\alpha \overset{\leftrightarrow}{\nabla}_\alpha  \Psi^{i}_\mu  -
m_\Delta \,  \Bar{\Psi}^{i}_\mu   \Psi^{i \mu}  -  g^{\lambda\sigma} \left( \Bar{\Psi}^{i}_\mu
i \gamma^{\mu}{\overset{\leftrightarrow}{\nabla}_\lambda}  \Psi^{i}_\sigma   +  \Bar{\Psi}^{i}_\lambda
i \gamma^{\mu}{\overset{\leftrightarrow}{\nabla}_\sigma}  \Psi^{i}_\mu  \right)  
\nonumber\\
&+&   i  \Bar{\Psi}^{i}_\mu \gamma^\mu \gamma^\alpha\gamma^\nu \overset{\leftrightarrow}{\nabla}_\alpha
\Psi^{i}_\nu + m_\Delta \Bar{\Psi}^{i}_\mu \gamma^\mu \gamma^\nu  \Psi^{i}_\nu 
\biggr\}\,,
\label{gdG}   
\\ 
S_{\Delta \rm N \pi}^{(1)} 
&=&
-\int d^4 x \sqrt{-g} ~ g_{\pi N \Delta} \bar \Psi \left( g^{\mu \nu} -\gamma^\mu\gamma^\nu\right) u_{\mu,i} \Psi_{\nu,i}  + \text{H.c.},
\label{gdnG} 
\end{eqnarray}
where the delta resonance is represented by the Rarita-Schwinger field\footnote{The delta
fields $\Psi^\mu_i$ contain isospin projectors $\xi_{ij}^{\frac{3}{2}}=\delta_{ij}-\frac{1}{3}\tau_{i}
\tau_{j}$, i.e. they satisfy the condition  $\Psi^\mu_i = \xi_{ij}^{\frac{3}{2}} \Psi^\mu_j.$},  $g^{\mu\nu}$ is the spacetime metric and $\gamma_\mu \equiv e_\mu^a \gamma_a $, with $ e_\mu^a$ denoting
the vielbein gravitational fields.~In Eqs. (\ref{gdG}) and (\ref{gdnG}) we take the off-shell parameter of the delta sector $A=-1$.~The building blocks of the above specified terms of the action are given in appendix \ref{A-I}.

   Further terms of the third and fourth order Lagrangians contributing to our calculations (at tree order) contain the Riemann tensor, the Ricci tensor and the Ricci scalar.~The action corresponding to the most general third- and fourth-order Lagrangian of such terms reduces to the following minimal form:
\begin{eqnarray}
S_{\rm \pi N}^{(2)}  & = & \int d^4x \sqrt{-g}\, \biggl\{\,  \frac{c_8}{8} R \bar\Psi\Psi + \frac{2 c_9}{m} R^{\mu \nu } \left( \bar\Psi \,  i  \gamma_\mu \overset{\leftrightarrow}{\nabla}_\nu \Psi\right)
\biggr\} \,, \label{PiNAction}
\\
S_{\rm \pi N}^{(3)} &=&  \int d^4x \sqrt{-g}\,\biggl\{\,  \tilde d_{g1}\,R \bar\Psi \gamma^\mu
\gamma_5 u_\mu \Psi +\tilde d_{g2}\,R^{\mu \nu} \bar\Psi \gamma_\mu \gamma_5 u_\nu \Psi + \tilde d_{g3} R^{\mu \nu \alpha \beta}  \left( \bar\Psi \sigma_{\mu \nu} \gamma_5 u_\alpha  \overset{\leftrightarrow}{\nabla}_\beta \Psi\right)
\nonumber
\\
&+&
 i \tilde d_{g4}  \nabla_\beta R^{\mu \nu \alpha \beta} \left( \bar\Psi \sigma_{\mu\nu}  \overset{\leftrightarrow}{\nabla}_\alpha \Psi\right)  +  \tilde d_{g5} \left(\bar\Psi \overset{\leftrightarrow}{\nabla}_\mu  \Psi \right) \partial^\mu R \biggr\},
 \label{newdterms}
 \\
 S_{\rm \pi}^{(4)} &=&  \int d^4x \sqrt{-g}\,\biggl\{\ l_{11} \, R \, {\rm Tr} ( D^\mu U  (D_\mu U)^\dagger )  + \ l_{12} \, R^{\mu \nu} \, {\rm Tr} ( D_\mu U  (D_\nu U)^\dagger ) + l_{13} \, R\,{\rm Tr}(\chi U^\dagger +U \chi^\dagger)  \biggr\}.
 \end{eqnarray}
The interactions with coupling constants $l_{11}, l_{12}$ and $l_{13}$ have been introduced in Ref.~\cite{Donoghue:1991qv} and the ones with $c_8$ and $c_9$  in  Ref.~\cite{Alharazin:2020yjv}.~Further we introduced in Eq.~(\ref{newdterms}) new interaction terms of the third order Lagrangian, with  coupling constants $\tilde d_{gi}$, contributing (at tree order) to our calculations.~Similarly to the terms with $c_8$, $c_9$, $l_{11}, l_{12}$ and $l_{13}$, the new interaction terms contribute in flat spacetime in the nucleon GFFs as well as in the OPGP although the corresponding action disappears in the flat spacetime.

\medskip

      The EMT for matter fields interacting with the gravitational metric field is given via
\begin{eqnarray}
T_{\mu\nu} (g,\psi) & = & \frac{2}{\sqrt{-g}}\frac{\delta S_{\rm m} }{\delta g^{\mu\nu}}\,,
\label{EMTMatter}
\end{eqnarray}
while for the fermion fields interacting with the gravitational vielbein fields we use 
\cite{Birrell:1982ix}:
\begin{eqnarray}
T_{\mu\nu}  (g,\psi) & = & \frac{1}{2 e} \left[ \frac{\delta S }{\delta e^{a \mu}} \,e^{a}_\nu
+ \frac{\delta S }{\delta e^{a \nu}} \,e^{a}_\mu  \right] \,.
\label{EMTfermion}
\end{eqnarray}

From the above specified terms of the action we obtain the EMT in flat spacetime.~The corresponding expressions are given in appendix \ref{A-I}.~In the next section we use the EMT to calculate the nucleon GFFs.   

    \section{One-Loop corrections to the gravitational form factors}
\label{II}

The one-nucleon matrix element of the EMT is parameterised in terms of three form factors as follows \cite{Polyakov:2018zvc}:
\begin{eqnarray}
\langle p_f, s_f| T_{\mu\nu}| p_i,s_i \rangle &=& \bar u(p_f,s_f) \left[ A(t) \frac{P_\mu P_\nu}{m_N} + i J(t) \frac{P_\mu \sigma_{\nu\alpha} \Delta^\alpha + P_\nu \sigma_{\mu\alpha} \Delta^\alpha}{2 m_N}+ D(t) \frac{\Delta_\mu \Delta_\nu-\eta_{\mu\nu} \Delta^2}{4 m_N} \right]  u(p_i,s_i) \,,
\label{EMTdef}
\end{eqnarray}
where $m_N$ is the mass of the nucleon, $(p_i,s_i)$ and $(p_f,s_f)$ are the momentum and polarization of the incoming and outgoing nucleons, respectively, and $P=(p_i+p_f)/2$, $\Delta=p_f-p_i$, $t = \Delta^2$.  

\begin{figure}[H]
	\centering
	\includegraphics[width=0.8 \textwidth]{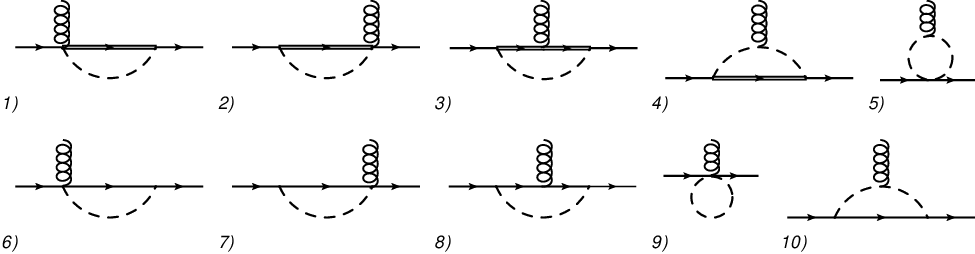}
	\caption{One-loop contributions to the nucleon GFFs.~Double, solid and dashed lines correspond to the deltas, nucleons and pions, respectively, while the curly lines represent gravitons.}
	\label{img:sp}
\end{figure}

Tree-order diagrams up to fourth order in the small scale expansion give the following contributions to the form factors:
\begin{eqnarray}
A_{\rm tree}(t) &=& 1 - \frac{2 c_9^\Delta}{m_N} \, t  + x_1^{\Delta} M_\pi^2\,  t  \,,\nonumber\\
J_{\rm tree}(t) &=& \frac{1}{2} - \frac{c_9^\Delta}{m_N} \, t + 2 m_N \tilde d_{g4}^\Delta t\,,\nonumber\\
D_{\rm tree}(t) &=&  c_8^\Delta m_N + y_1^{\Delta}\,  t + y_2^{\Delta} M_\pi^2 \,,
\label{temt}
\end{eqnarray} 
where $x_1^{\Delta}$ and $y_i^{\Delta}$ parametrize the tree-order contributions of the fourth chiral order.~We use the superscript $\Delta$ to distinguish the LECs from the ones, that appear in the theory without $\Delta$ resonances.~For the one-loop contributions we calculate the diagrams shown in Fig.~\ref{img:sp}.

To order the contributions of various diagrams we apply the small scale 
expansion (also called the $\epsilon$-counting scheme) \cite{Hemmert:1996xg,Hemmert:1997ye}\footnote{For an alternative power counting in an EFT with delta resonances, see Ref.~\cite{Pascalutsa:2002pi}.}.
Within this scheme the pion lines count as of order minus two, the nucleon and delta lines
have order minus one, interaction vertices originating from the effective Lagrangian 
of order $N$ count also as of order $N$ and the vertices generated by the EMT
have the orders corresponding 
to the number of quark mass factors and derivatives acting on the pion fields. 
Derivatives acting on the nucleon and delta fields count as of order 
zero. The momentum transfer between the initial and final states counts as of order one, 
therefore in those terms of the EMT which contain full derivatives, these derivatives
count as of order one.  
Integration over loop momenta is counted as of order four. The delta-nucleon
mass difference also counts as of order one within the $\epsilon$-counting scheme.~It is understood that the above described power counting for loop diagrams is realized as the result 
of our manifestly Lorentz-invariant calculations only after performing an appropriate
renormalization.  To get rid of the divergent parts and power counting violating pieces
from the expressions of one-loop diagrams we apply the  EOMS scheme of
Refs.~\cite{Gegelia:1999gf,Fuchs:2003qc}, with the remaining renormalization scale chosen as $\mu = m_N$, where $m_N$ is the mass of the nucleon.~The one-loop parts of counter terms are given by:
\begin{eqnarray}
\delta c_8^\Delta & = &\frac{1}{F_\pi^2} \left[ \frac{16 g_{\pi N \Delta}^2 }{27 } (25 \delta +9 m_N) + g_A^2 m_N ( 9 c_8^{\Delta R} m_N + 48 c_9^{\Delta R} m_N - 4) \right] \lambda_b + \frac{g_A^2 m_N (3 c_8^{\Delta R} m_N +1)}{16 \pi^2 F_\pi^2} 
\nonumber
\\
&+& \frac{5 g_{\pi N \Delta}^2 c_8^{\Delta R} m_N}{3 F_\pi^2}(6 \delta + m_N) \lambda_b + \frac{ g_{\pi N \Delta}^2 c_8^{\Delta R} m_N}{576 \pi^2 F_\pi^2} (7 m_N - 78 \delta)  -\frac{ g_{\pi N \Delta}^2 m_N}{48 \pi^2 F_\pi^2} 
 \, ,
\nonumber
\\
\delta c_9^\Delta &=& \frac{  g_{\pi N \Delta}^2}{ 27 F_\pi^2} \left[ 45 c_9^{\Delta R} m_N (6 \delta + m_N) - 4 (\delta + 9 m_N) \right] \lambda_b + \frac{131 g_{\pi N \Delta}^2~ m_N}{3456 \pi^2 F_\pi^2} + \frac{g_{\pi N \Delta}^2 c_9^{\Delta R} m_N}{576 F_\pi^2 \pi^2} (7 m_N -78 \delta)\nonumber
\\
&+&
\frac{3 g_A^2 m_N}{128 \pi^2 F_\pi^2} \, ,
\nonumber
\\
\delta  \tilde d_{g4}  &=&   \frac{g_{\pi N \Delta}^2 }{54 F_\pi^2~ m_N^2} (23  m_N - 4 \delta) \lambda_b + \frac{131 g_{\pi N \Delta}^2}{6912 \pi^2 F_\pi^2~ m_N}\, ,
\label{counterterms}
\end{eqnarray}
where
\begin{eqnarray}
\lambda_b &=&  \frac{1}{32 \pi^2} \left(  \frac{2}{n-4} -1 + \gamma - \ln(4\pi)\right),\,
\end{eqnarray}
$n$ is the spacetime dimension, $\gamma=-\Gamma'(1)$ and $\delta$ is the delta-nucleon mass difference, i.e. $\delta = m_{\Delta} - m_N$.

    After renormalization we obtain for the $D-$term the following one-loop contribution, expanded in powers of the pion mass 
and the delta-nucleon mass difference:
\begin{eqnarray}
D_{\rm loop}(0) 
&=& -\frac{ g_{\pi N \Delta}^2~ m_N}{54 F_\pi^2 \pi^2 } \left[ 17 \delta -12 \left[  \sqrt{\delta^2 - M_\pi^2} - \delta \right] \ln \left( \frac{M_\pi}{m_N}\right)  +12  \sqrt{\delta^2 - M_\pi^2} \ln \left( \frac{ \delta +  \sqrt{\delta^2 - M_\pi^2}  }{m_N}\right)  \right]
\nonumber
\\
&-&
\frac{g_{\pi N \Delta}^2}{1728 \pi^2 F_\pi^2} \left[ \delta^2 (555 c_8^\Delta m_N +392) + 24 M_\pi^2 (18 c_8^\Delta m_N +11)
\right] 
\nonumber
\\
&-&
\frac{g_{\pi N \Delta}^2}{18 \pi^2 F_\pi^2} \left[ M_\pi^2 (9 c_8^\Delta m_N + 3 ) + 2 \delta (9 c_8^\Delta m_N + 7) \left[  \sqrt{\delta^2 - M_\pi^2} - \delta \right]
\right]   \ln \left( \frac{M_\pi}{m_N}\right)
\nonumber
\\
&+&
\frac{g_{\pi N \Delta}^2 \delta (9 c_8^\Delta m_N + 7) }{9 \pi^2 F_\pi^2}  \sqrt{\delta^2 - M_\pi^2} \ln \left( \frac{ \delta +  \sqrt{\delta^2 - M_\pi^2}  }{m_N}\right) 
\nonumber
\\
&+&
\frac{ g_A^2 m_N}{16 \pi  F_\pi^2} \,M_{\pi } 
+\frac{ -3 g_A^2 +2 m_N\left(-4
   c_1^\Delta+c_2^\Delta+2 c_3^\Delta\right) }{8 \pi ^2 F_\pi^2} \, M_{\pi }^2 \ln\left( \frac{M_{\pi }}{m_N }\right) \nonumber\\
&+& \frac{\left(- g_A^2 \left(3 c_8^\Delta m_N + 14 \right)+8 c_3^\Delta m_N-16 c_1^\Delta m_N\right)}{32 \pi ^2
   F_\pi^2}\ M_{\pi }^2 
    + \mathcal{O} (\epsilon^3)\, .
\nonumber
\\
\label{FFs0}
\end{eqnarray}

\begin{figure}[H]
\begin{minipage}[h]{0.495\linewidth}
\center{\includegraphics[width=0.75\linewidth]{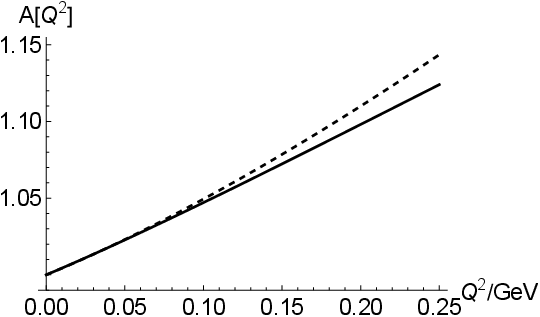}} \\
\end{minipage}
\hfill
\begin{minipage}[h]{0.495\linewidth}
\center{\includegraphics[width=0.75\linewidth]{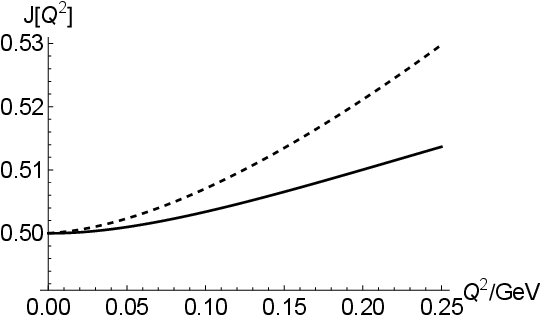}} \\
\end{minipage}
\vfill
\vspace{0.5cm}
\center{\includegraphics[width=0.393\linewidth]{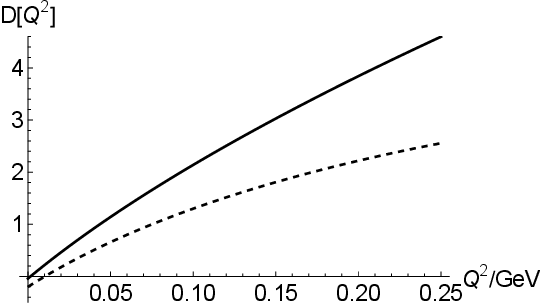}} \\
\caption{GFFs of the nucleon as functions of $Q^2$. Solid curves represent the GFFs with $\Delta$ resonances while dashed curves stand for GFFs without $\Delta$ resonances.}
	\label{NGFF:sp}
\end{figure}

Next, we define the slopes of the GFFs by writing the form factors as:
\begin{eqnarray}
A(t) & = & 1 + s_{A} t +{\cal O}(t^2) \,,
\nonumber
\\
J(t) & = & \frac{1}{2} + s_{J} t +{\cal O}(t^2) \,,
\nonumber
\\
D(t) & = & D(0) + s_{D} t +{\cal O}(t^2) \,.
\label{defradii}
\end{eqnarray}
Calculating loop contributions to these quantities and expanding in powers of the pion mass and $\delta$ we obtain
\begin{eqnarray}
s_{A_{\rm loop}} 
&=&
\frac{g_{\pi N \Delta}^2}{432 F_\pi^2 \pi^2 m_N (M_\pi^2-\delta^2) } \biggl[ 
7 \delta (M_\pi^2 - \delta^2) -12 \left[ 12 \delta^2 \left( \delta -\sqrt{\delta^2 - M_\pi^2}  \right)  - M_\pi^2 \left( 12 \delta - 7 \sqrt{\delta^2 - M_\pi^2} \right)  \right]   \ln \left( \frac{M_\pi}{m_N}\right)
\nonumber
\\
&+&
12 (7 M_\pi^2 - 12 \delta^2) \sqrt{ \delta^2 - M_\pi^2 }   \ln \left( \frac{ \delta +  \sqrt{\delta^2 - M_\pi^2}  }{m_N}\right) \biggl]  + \frac{g_{\pi N \Delta}^2}{ 864 F_\pi^2 \pi^2 m_N^2} \left[ \delta^2 (555 c_9^\Delta m_N - 257) \right.
\nonumber
\\
& +&
\left.
 M_\pi^2 (432 c_9^\Delta m_N -43) \right] + \frac{g_{\pi N \Delta}^2 \delta}{72 \pi^2 F_\pi^2 m_N^2 \sqrt{\delta^2 - M_\pi^2} } \biggl[ 4 \delta^2 (11 - 36 c_9^\Delta m_N) +3 M_\pi^2 ( 48 c_9^\Delta m_N - 13 ) \biggl]    \ln \left( \frac{ \delta +  \sqrt{\delta^2 - M_\pi^2}  }{m_N}\right)
\nonumber
\\
&+&
 \frac{g_{\pi N \Delta}^2 }{72 \pi^2 F_\pi^2 m_N^2 ( M_\pi^2-\delta^2) } \biggl[ 3 M_\pi^4 (24 c_9^\Delta m_N -7)+ \delta M_\pi^2 \left( \delta (65 -216 c_9^\Delta m_N) + 3 (48 c_9^\Delta m_N - 13) \sqrt{\delta^2 - M_\pi^2} \right) \nonumber
\\
&-&
4 \delta^3 (36 c_9^\Delta m_N -11) \left( \sqrt{\delta^2 - M_\pi^2} -\delta \right) \biggl] \ln \left( \frac{M_\pi}{m_N}\right) -\frac{7  g_A^2}{128 \pi  F_\pi^2 m_N}\ M_{\pi } + \frac{ \left(c_2^\Delta m_N-4 g_A^2\right)}{16 \pi ^2 F_\pi^2 m_N^2}\ M_{\pi }^2 \ln \left(\frac{M_{\pi }}{m_N}\right)
\nonumber
\\
&-&
\frac{3
    g_A^2 \left(2 c_9^\Delta m_N+1\right)}{32 \pi ^2 F_\pi^2 m_N^2}\ M_{\pi }^2  + \mathcal{O} (\epsilon^3)\,,
\nonumber
\\
s_{J_{\rm loop}} 
&=& \frac{g_{\pi N \Delta}^2}{1152 F_\pi^2 \pi^2 (M_\pi^2-\delta^2) } \biggl[ 3 (M_\pi^2 - \delta^2) + 32 \left[ \delta \left( \sqrt{\delta^2 - M_\pi^2} - \delta \right) + M_\pi^2 \right]  \ln \left( \frac{M_\pi}{m_N}\right)
\nonumber
\\
&-&
32 \delta \sqrt{\delta^2 - M_\pi^2 }  \ln \left( \frac{ \delta +  \sqrt{\delta^2 - M_\pi^2}  }{m_N}\right) \biggl] - \frac{13 g_{\pi N \Delta}^2 \delta }{64 F_\pi^2 \pi^2 m_N } + \frac{g_{\pi N \Delta}^2 (12 \delta^2 - 7 M_\pi^2) }{72  F_\pi^2 \pi^2 m_N  \sqrt{\delta^2 - M_\pi^2} }  \ln \left( \frac{ \delta +  \sqrt{\delta^2 - M_\pi^2}  }{m_N}\right) 
\nonumber
\\
&+&\frac{g_{\pi N \Delta}^2 }{72  F_\pi^2 \pi^2 m_N  (M_\pi^2- \delta^2 ) } \left[ M_\pi^2 \left( 12 \delta - 7 \sqrt{\delta^2 - M_\pi^2} \right)  + 12 \delta^2  \left( \sqrt{\delta^2 - M_\pi^2} - \delta \right) \right]  \ln \left( \frac{M_\pi}{m_N}\right)
\nonumber
\\
&+& \frac{g_A^2 \left(4 c_9^\Delta m_N-5\right)}{64 \pi ^2 F_\pi^2} 
  +\frac{7  g_A^2}{128 \pi  F_\pi^2 m_N}\ M_{\pi }  
  + \mathcal{O} (\epsilon^2)\,,
\nonumber
\\
s_{D_{\rm loop}} 
&=&
\frac{4g_{\pi N \Delta}^2 m_N }{45 F_\pi^2 \pi^2 \sqrt{\delta^2 -M_\pi^2} } \left[  \ln \left( \frac{M_\pi}{m_N}\right) -  \ln \left( \frac{ \delta +  \sqrt{\delta^2 - M_\pi^2}  }{m_N}\right) \right] 
\nonumber
\\
&-&\frac{g_A^2 m_N}{40 \pi  F_\pi^2 }\ \frac{1}{M_\pi} -\frac{ \left(5 g_A^2+4 \left(c_2+5 c_3\right) m_N\right)}{80 \pi ^2
   F_\pi^2} \ \ln \left(\frac{M_{\pi }}{m_N}\right)+ { \frac{g_A^2 \left(24+ (15 c_8+40 c_9^\Delta) m_N \right)}{480 \pi ^2 F_\pi^2} } 
\nonumber
\\
&+&
\frac{g_{\pi N \Delta}^2}{12960 F_\pi^2 \pi^2 (M_\pi^2-\delta^2) } \biggl[ 
461 (M_\pi^2 - \delta^2) -144 \left[ \delta \left( 3 \sqrt{\delta^2 - M_\pi^2} - 19 \delta \right) + 19 M_\pi^2 \right]   \ln \left( \frac{M_\pi}{m_N}\right)
\nonumber
\\
&+&
432 \delta \sqrt{ \delta^2 - M_\pi^2 }   \ln \left( \frac{ \delta +  \sqrt{\delta^2 - M_\pi^2}  }{m_N}\right) \biggl] +
\frac{\left(4 c_1^\Delta-c_2^\Delta-7 c_3^\Delta\right) m_N}{40 \pi ^2 F_\pi^2} + \mathcal{O} (\epsilon)\,.
\label{radii}
\end{eqnarray}
As mentioned in the introduction, the one-loop corrections to the GFFs of the nucleon were calculated in Ref.~\cite{Alharazin:2020yjv} in the framework of chiral EFT.~If we switch off the delta resonances, i.e. set $g_{\pi N \Delta} = 0$, then we obtain the same expressions for the slopes and the D-term as the ones obtained in  Ref.~\cite{Alharazin:2020yjv}.~To illustrate the differences between GFFs with and without delta contributions we use the programm $LoopTools$ \cite{Hahn:1999mt} and plot the GFFs as functions of $Q^2 = -t$ in Fig.~\ref{NGFF:sp}.~Before doing that we had to fix the LECs $c_8, c_9$ and $\tilde d_{g4}$ together with the parameters $x_1$ and $y_i$ in both theories.~Due to the lack of empirical data we cannot fix all of them, but for the sake of comparison we substitute the following values of the unknown LECs in the theory without $\Delta$ resonances
 \begin{eqnarray}
x_1 = y_1= y_2 = \tilde d_{g4} = c_9 = 0\,, 
 \label{unknown}
 \end{eqnarray}
and fix $c_8$ from the value of the D-term and the LECs of the effective Lagrangian with explicit delta from the condition of matching physical quantities in theories with and without explicit deltas:
 \begin{eqnarray}
D^\Delta &=& D = -0.2\, ,
\nonumber
\\
s_{A}^\Delta &=& s_{A}
\, ,
\nonumber
\\ 
s_{J}^\Delta &=& s_{J}
\, ,
\nonumber
\\     
s_{D}^\Delta &=& s_{D}
\,,
\label{Fixing-1}
\label{conditions5}
\end{eqnarray}
 where $D^\Delta $ and $s_{i}^\Delta $ correspond to the D-term and the slopes in the theory with $\Delta$ resonances.~The value $D = - 0.2$ is taken from Ref.~\cite{Gegelia:2021wnj}\footnote{The authors of Ref.~\cite{Gegelia:2021wnj} obtained for a model independent bound $D \leq -0.2$.~For the sake of the current work we take the value $ D=-0.2$ to fix the coupling constant $c_8$.}.~For the choice of the unknown LECs given in Eq.~(\ref{unknown}) we obtain from  Eq.~(\ref{Fixing-1}) the following values for the remaining parameters
\begin{eqnarray}
&&x_1^\Delta = 1 ~\text{GeV}^{-4}, \quad y_1^\Delta = -7.2~\text{GeV}^{-2}, \quad y_2^\Delta = 12.2~\text{GeV}^{-2}, \quad  c^\Delta_8 = - 0.97~\text{GeV}^{-1}, 
\nonumber
\\
&&\quad c_8 = -1.15~\text{GeV}^{-1} ,  \quad  c^\Delta_9 = -1.29~\text{GeV}^{-1}, \quad \tilde d_{g4}^\Delta = 0.44~\text{GeV}^{-3}.
\end{eqnarray} 
Moreover, we use for the plots below the following numerical values for the masses and the LECs obtained in Ref.~\cite{Siemens:2016hdi}
\begin{eqnarray}
&& g_{\pi N \Delta}= 1.35,\quad g_A = 1.289,\quad  m_N = 0.93827~\text{GeV} ,\quad m_\Delta = 1.232~\text{GeV}, \quad  M_\pi = 0.13957~\text{GeV},
\nonumber
\\
&& F_\pi = 0.0922~\text{GeV},\quad c_1= -0.82 ~\text{GeV}^{-1},\quad c_2= 3.56 ~\text{GeV}^{-1},\quad c_3 = -4.59 ~\text{GeV}^{-1},
\nonumber
\\
&&  c_1^\Delta= -1.15 ~\text{GeV}^{-1}, \quad  c_2^\Delta= 1.57 ~\text{GeV}^{-1},\quad c_3^\Delta = -2.54 ~\text{GeV}^{-1}.
\end{eqnarray}
As can be seen in Fig.~\ref{NGFF:sp} the contributions with the $\Delta$ resonances become important for the GFFs with increasing $Q^2$ (for more details see below).~As the next step, we calculate the long-range behavior of the corresponding gravitational spatial densities, which can be obtained from the non-analytic contributions of the GFFs in the chiral limit.~That is, we decompose the GFFs in two parts, one which is analytic in $t$ and one which is non-analytic\footnote{The analytic contributions are too lengthy to be shown here and they are not relevant for coming discussion.}.~Similarly to Ref.~\cite{Alharazin:2020yjv} we expand the non-analytic contributions of the GFFs in the chiral limit and obtain up to the accuracy of our calculations the following results:

\begin{eqnarray}
A (t) &=&\frac{g_{\pi N \Delta}^2}{\pi^2 F_\pi^2 m_N^2} \left(\frac{(79 \delta + 10 m_N)}{5760 \delta}t^2 + \frac{(15 \delta + 2 m_N)}{2304 \delta^3}t^3 \right)  \ln\left( -\frac{t}{m_N^2}\right) + \frac{3 g_A^2  }{512 F_\pi^2 m_N} (-t)^{\frac 32} \nonumber
\\
&-&\frac{ \left(c_2 m_N-10 g_A^2\right)}{320 \pi ^2 F_\pi^2 m_N^2} \ t^2 \ln\left( -\frac{t}{m_N^2}\right) +
 {   \mathcal{O} (t^{\frac{5}{2}}) }  \,, 
\nonumber
\\
J (t) &=& \frac{g_{\pi N \Delta}^2 \left(4 \delta (2 \delta +m_N)t^2 + t^3 \right) }{2304 F_\pi^2 \pi^2 \delta^3 m_N} \ln \left(-\frac{t}{m_N^2} \right)  -\frac{ g_A^2 }{64 \pi ^2 F_\pi^2} t \ln \left(-\frac{t}{m_N^2}\right)   -\frac{3 g_A^2  }{512 F_\pi^2 m_N} 
   (-t)^{\frac 32} + \mathcal{O} (t^2)  
    \,,\nonumber\\
D(t) &=& - \frac{g_{\pi N \Delta}^2  ( 214 \delta^3 + 10 m_N ( t -14 \delta^2) + 5 \delta t) }{2880 F_\pi^2 \pi^2 \delta^3} t \ln \left(-\frac{t}{m_N^2} \right) +\frac{3
    g_A^2 m_N}{128 F_\pi^2}\sqrt{-t}  \nonumber
    \\
    &-&
    \frac{\left(5 g_A^2+4 \left(c_2+5 c_3\right) m_N\right)}{160 \pi ^2 F_\pi^2}\  t
   \ln \left(\frac{ - t}{m_N^2} \right) + { \mathcal{O} (t^{\frac{3}{2}})} \,.
   \label{asymptoticsFF}
   \end{eqnarray}

\begin{figure}[H]
\begin{minipage}[h]{0.495\linewidth}
\center{\includegraphics[width=0.75\linewidth]{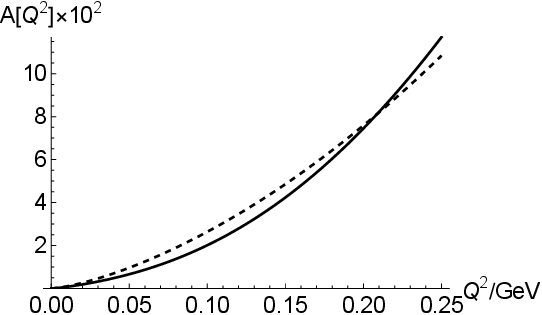}} \\
\end{minipage}
\hfill
\begin{minipage}[h]{0.495\linewidth}
\center{\includegraphics[width=0.75\linewidth]{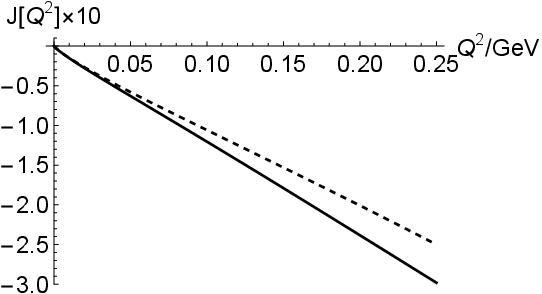}} \\
\end{minipage}
\vfill
\vspace{0.5cm}
\center{\includegraphics[width=0.393\linewidth]{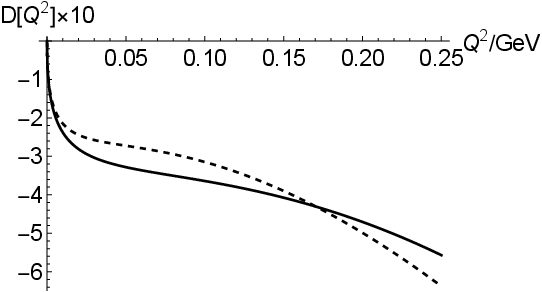}} \\
\caption{Non-analytic contributions of the GFFs of the nucleon as functions of $Q^2$. Solid curves represent the GFFs with $\Delta$ resonances while dashed curves stand for GFFs without $\Delta$ resonances.}
	\label{GFF-Non-Analytic}
\end{figure}

For the sake of comparison we plot in Fig.~\ref{GFF-Non-Analytic} the non-analytic parts of GFFs with and without delta contributions.
As we can see from Fig.~\ref{GFF-Non-Analytic}, the $\Delta$ resonances give noticeable contributions to the non-analytic parts of the GFFs, which in its turn will affect the spatial density distributions directly.~The importance of these effects will be discussed at the end of this section.

   The corresponding connections between the GFFs and the spatial densities of energy ($\rho_E$), spin ($\rho_J$), pressure ($p$) and shear forces ($s$)  in the zero average momentum frame (ZAMF) are given in \cite{Panteleeva:2023evj} and they read 
 \begin{eqnarray}
\rho_E (r) &=& N_{\phi, R} \int \frac{d^2 n d^3 q}{(2\pi)^3} A(-{\bf q}^2_{\perp})e^{- i  {\bf r}{\cdot} {\bf q} }\, ,
\nonumber
\\
\rho_J (r)&=& -  \frac{N_{\phi, R}}{2 m_N}  ~r  \frac{d}{d r} \int \frac{d^2 n d^3 q}{(2\pi)^3} J(-{\bf q}^2_{\perp})e^{- i  {\bf r}{\cdot} {\bf q} }\, ,
\nonumber
\\
s(r) &=& N_{\phi, R, 2} \int \frac{ d^2  \hat n d^3 q}{(2\pi)^3} \left( \frac{3 ( {\bf r}{\cdot} {\bf q})^2 - {\bf r}^2{\bf q}^2 }{4 {\bf r}^2 } \right) D(-{\bf q}^2_{\perp} ) e^{- i  {\bf r}{\cdot} {\bf q} }\, ,
\nonumber
\\
p(r) &=& \frac{ N_{\phi, R, 2}}{2} \int \frac{ d^2  \hat n d^3 q}{(2\pi)^3} \left( \frac{{\bf q}^2}{3} -{\bf q}^2_{\perp} \right) D(-{\bf q}^2_{\perp} ) e^{- i  {\bf r}{\cdot} {\bf q} }\,.
\label{Spatial-Densities}
\end{eqnarray}
Using the expressions of Eqs.~(\ref{asymptoticsFF}) and (\ref{Spatial-Densities}) we obtain the following results (valid in the region of distances $1/\Lambda_{\text{strong}} \ll r \ll 1/M_\pi$):
\begin{eqnarray}
\rho_E (r) &=&   N_{\phi, R}  \left( \frac{27 g_A^2}{512 F_\pi^2 m_N} \frac{1}{r^6}  - \frac{g_{\pi N \Delta}^2 (79 \delta + 10 m_N)}{45\pi^2 F_\pi^2 m_N^2\delta}\frac{1}{r^7} + \frac{ 2 (c_2 m_N - 10 g_A^2)}{5\pi^2 F_\pi^2 m_N^2 }\frac{1}{r^7}   \right)+ \mathcal{O} \left(\frac{1}{r^8} \right)\,,
\nonumber 
\\
\rho_J (r)&=&  N_{\phi, R} \left(  \frac{5 g_A^2}{16 \pi^2  m_N F_\pi^2} \frac{1}{r^5}  - \frac{81 g_A^2}{512 F_\pi^2 m_N^2} \frac{1}{r^6} -  \frac{7 g_{\pi N \Delta}^2 (2 \delta +m_N) }{9 F_\pi^2 \pi^2 \delta^2 m_N^2} \frac{1}{r^7} \right)+ \mathcal{O} \left(\frac{1}{r^8} \right)\,,
\nonumber
\\
s(r) &=& N_{\phi, R, 2}  \left( \frac{9 g_A^2 m_N}{32 F_\pi^2} \frac{1}{r^6} +\frac{7 g_{\pi N \Delta}^2 (70 m_N - 107 \delta )}{72 \pi^2 F_\pi^2 \delta } \frac{1}{r^7} -\frac{7 \left(5 g_A^2+4 \left(c_2+5 c_3\right) m_N\right)}{ 8 \pi^2 F_\pi^2}  \frac{1}{r^7} \right) + \mathcal{O} \left(\frac{1}{r^8} \right)\,,
\nonumber
\\
p(r) &=&  - N_{\phi, R, 2} \left(  \frac{15 g_A^2 m_N}{256 F_\pi^2}\frac{1}{r^6} + \frac{7 g_{\pi N \Delta}^2 (70 m_N -107 \delta) }{ 270 \pi^2 F_\pi^2 \delta } \frac{1 }{ r^7}  - \frac{7 \left(5 g_A^2+4 \left(c_2+5 c_3\right) m_N\right) }{30 \pi ^2 F_\pi^2}\frac{1 }{ r^7}  \right)  + \mathcal{O} \left(\frac{1}{r^8} \right)\,.
\end{eqnarray}
The local stability condition of Ref.~\cite{Polyakov:2018zvc} (i.e. $\frac{2}{3} s(r) + p(r) \geq 0$) is satisfied by all terms, while the positivity of the energy density is satisfied by all terms except the ones that are proportional to $g_{\pi N \Delta}^2$.~These terms that violate the positivity of $\rho_E (r)$ are related to the decay of delta to a nucleon and a pion, i.e. they carry information about the instability of the system.~This behavior is similar to the one obtained in Ref.~\cite{Epelbaum:2021ahi}, where the spatial densities of the $\rho$ meson were calculated and it was shown, that the general stability condition and the positivity of the energy density are satisfied by all terms except the ones that depend on the coupling constant corresponding to the decay of the $\rho$ meson.

\subsection{Importance of the $\Delta$ resonances to nucleon GFFs}
Below we estimate the importance of including the $\Delta$ resonances as explicit degrees of freedom when dealing with the GFFs of the nucleon.~Generally in effective field theories we expect the coupling constants to be dominated by masses of the lightest particles, that have been integrated out.~In the framework of considered chiral EFT, we expect that coupling constants are scaled with the $\rho$ meson mass ($M_\rho = 0.75\, \text{GeV}$ \cite{ParticleDataGroup:2020ssz}).~Moreover, when counting the orders of small parameters, we mean implicitly expansion in massless scales, which are given in our case by  $ \frac{M_\pi}{M_\rho},   \frac{\delta}{M_\rho}$ and  $\frac{Q^2}{M_\rho^2} $.~Since we are doing the calculations up to fourth chiral order, the error of our calculations is expected to be of order five, which means, that for $Q^2 \sim M_\pi^2 \approx 0.019\, \text{GeV}^2$ the error due to higher chiral order contributions is expected to be around $0.02\%$.~From the full results we have at $Q^2=M_\pi^2$ the following values for the GFFs with and without $\Delta$ contributions
\begin{eqnarray}
A^\Delta(M_\pi^2) &=& 1.0086\,, \quad A(M_\pi^2) = 1.0085\,, 
\nonumber
\\
J^\Delta(M_\pi^2) &=& 0.5001 \,, \quad J(M_\pi^2) =0.5005\,, 
\nonumber
\\
D^\Delta(M_\pi^2) &=& 0.45\,, \quad\,\,\,\, D(M_\pi^2) = 0.17\,.~
\label{num-re}
\end{eqnarray} 
From Eq.~(\ref{num-re}) we have differences of $0.01\%, 0.08\%$ and $62\%$ for $A(t), J(t)$ and $D(t)$, respectively.~Except for $A(t)$, these differences lie beyond the estimated error at $t = M_\pi^2$ and hence, we conclude that the inclusion of the $\Delta$ resonances is important for the GFFs of the nucleon.~As we can see from the plotted full results, the $\Delta$ resonances give larger contributions to $D(t)$ than to $A(t)$ and $J(t)$.~This can be explained as follows: The structure of EMT is such that, at $t = 0$ we have in both theories $A(0) = 1$ which corresponds to mass of the nucleon and $J(0) = \frac{1}{2}$ which corresponds to the spin, while for $D(0)$ the value is not fixed by the structure of the EMT.~Hence, for $D(t)$ we expect more deviation between both theories around $t=0$, i.e. also at $ t= M_\pi^2$ than for $A(t)$ and $J(t)$.~It is worth mentioning, that while the above numerical values correspond to the choice of Eq.~(\ref{unknown}), we obtained that for various natural values of the unknown LECs the resulting GFFs behave similarly to Fig.~\ref{NGFF:sp}.~Moreover, we can support the observation, that the contributions of the delta resonances are important by considering the non-analytic parts of the GFFs (see Eq.~(\ref{asymptoticsFF}) and Fig.~\ref{GFF-Non-Analytic}), for which all LECs are known and fixed via experiments.~At $Q^2=M_\pi^2$ we have the following values:
\begin{eqnarray}
A^\Delta(M_\pi^2) &=&  1.8 ~{\cdot} 10^{-3}\,,\quad  \,\, A(M_\pi^2) = 2.5  ~{\cdot} 10^{-3}\,, 
\nonumber
\\
J^\Delta(M_\pi^2) &=& -27  ~{\cdot}10^{-3}\,,\quad J(M_\pi^2) =-26   ~{\cdot} 10^{-3}\,\,, 
\nonumber
\\
D^\Delta(M_\pi^2) &=& -0.28\,, \quad \,\,\,\,\,\,\,\,\,\, D(M_\pi^2) = -0.24\,.~
\label{non-num-re}
\end{eqnarray} 

From the above results we have differences of $28\%, 3.7\%$ and $14.3\%$ for $A(t), J(t)$ and $D(t)$, respectively.~All of these differences lie beyond the estimated error of $0.02\%$, due to higher chiral orders.

To improve the above calculations one might consider higher order contributions coming from one-loop corrections that contain vertices from the action given in Eq.~(\ref{newdterms}).~However, such contributions contain many unknown LECs.~This suggests to study various processes of gravity-induced hadronic interactions, in which the unknown LECs contribute and hence can be determined.~An example for such a process is the transition of the nucleon to the nucleon and the pion via the gravitational field, whose matrix element and Lorentz structure is studied in the following section.

\section{Matrix element of the one pion graviproduction}
 \label{Sec:2}
 In this section we discuss the OPGP off the nucleon, i.e. the matrix element of the EMT where the initial state contains one nucleon and the final state contains one nucleon and one pion.~As mentioned in the introduction the OPGP, which is accessible in hard exclusive processes like the non-diagonal DVCS ($\gamma^* + N \mapsto \gamma + (\pi N) $) can be used to obtain additional information about the new LECs of chiral EFT \cite{Max:1,Guichon:2003ah,Max:2}, which are needed to make predictions for other physical processes.~The measurement of the amplitude of the OPGP is planned by the CLAS12 collaboration at JLab (USA) \cite{Alharazin:2020yjv}.

    The matrix element corresponding to OPGP has a lengthy Lorentz structure, as can be seen in Ref.~\cite{Max:2}.~This leads to technical complications in calculations, like checking the current conservation.~Therefore it would be helpful to parametrize the matrix element of OPGP in terms of independent, conserved Lorentz invariant structures.~This will be done below.

    The amplitude of the OPGP has the following general form 
\begin{eqnarray}
-i \left<p_f,s_f, q, c \right| T^{\mu \nu}  \left| p_i,s_i \right>  =  \bar u(p_f,s_f) O^{\mu \nu} \tau^c u(p_i,s_i),
\end{eqnarray}
where $q$ is the momentum of the pion with isospin index $c$, $p_i$ and $p_f$ are the momenta of the incoming and outgoing nucleons, respectively, and $O^{\mu \nu}$ is a symmetric matrix in Dirac space.~In the following we make use of the Lorentz invariance and conservation of the EMT to  parametrize $O^{\mu \nu}$ in terms of independent and separately conserved Lorentz structures.~For that sake we introduce first the following linearly independent kinematic variables
\begin{eqnarray}
\tilde \Delta &=& p_f + q - p_i  ,\\
P &=& p_f - q + a~ p_i ,\\  
\Lambda &=& p_f + q  + b~ p_i, 
\end{eqnarray}
where
\begin{eqnarray}
a &=& -1 -  \frac{2 M_\pi^2  -2 t }{m_N^2 - s + \tilde t}\,,
\nonumber
\\
b &=&\frac{-m_N^2 + s + \tilde t}{m_N^2 - s + \tilde t}\,,
\nonumber 
\end{eqnarray} 
and  $s =\left( p_f +q \right)^2$, $t = \left( p_f-  p_i \right)^2$ and $\tilde t = \tilde \Delta^2$.~One can easily check that $\tilde \Delta {\cdot} P =0, ~\tilde \Delta {\cdot} \Lambda =0$.
Analogously to Ref. \cite{Cotogno:2019vjb}, applying the Cayley-Hamilton theorem, we construct $O^{\mu \nu}$ for spin-1/2 systems using two different covariant multipoles contracted with the kinematic variables in all possible ways that respect symmetry and conservation of the EMT.~Moreover, since the pion field has negative parity, it gives an overall minus sign under parity transformation and hence the tensor $O^{\mu \nu}$ must transform with minus 
under parity.
 \footnote{This is because the parity is conserved for the matrix element of the EMT.}~This means that each structure in  $O^{\mu \nu}$ must contain either $\epsilon^{\alpha\beta\gamma\kappa}$ or $\gamma_5$.~We make use of the Lorentz invariance, symmetry and the conservation of the EMT to obtain all possible structures contributing to $O^{\mu \nu}$.~After investigating the dependence of these structures on each other and removing the redundant ones, we find that the most general form of the matrix element contains twelve independent structures and it can be written as
\begin{eqnarray}
-i \left<p_f,s_f, q, c \right| T^{\mu \nu}  \left| p_i,s_i \right>  &=& \bar u(p_f,s_f) \Biggl[ \frac{1}{2} \left(f_1 \tilde t \gamma^5 +  \frac{i f_{2}}{m_N} \epsilon^{\tilde \Delta P \Lambda \beta} \gamma_\beta \right)P^\mu P^\nu  +  \frac{1}{2} \left(f_3 \tilde t \gamma^5 +  \frac{i f_{4}}{m_N} \epsilon^{\tilde \Delta P \Lambda \beta} \gamma_\beta \right)\Lambda^\mu \Lambda^\nu 
\nonumber
\\
&+&  \frac{1}{2} \left(f_5 \tilde t \gamma^5 +  \frac{i f_{6}}{m_N} \epsilon^{\tilde \Delta P \Lambda \beta} \gamma_\beta \right) P^\mu \Lambda^\nu +\frac{ 1}{2} \left( f_{7} \tilde t \gamma^5+  \frac{i f_{8}}{m_N} \epsilon^{\tilde \Delta P \Lambda \beta} \gamma_\beta  \right)\left( \tilde t \eta^{\mu \nu} -\tilde \Delta^\mu \tilde \Delta^\nu\right)
\nonumber
\\
&+& i f_9 \left( \tilde t \epsilon^{\nu P \Lambda \beta}- \epsilon^{\tilde \Delta P \Lambda \beta}\tilde \Delta^\nu \right) P^\mu\gamma_\beta + i f_{10} \left( \tilde t \epsilon^{\nu P \Lambda \beta} - \epsilon^{\tilde \Delta P \Lambda \beta}\tilde \Delta^\nu \right)\Lambda^\mu\gamma_\beta
\nonumber\\
&+& 
 i \left( f_{11} ~  \Lambda^\mu + f_{12} ~  P^\mu  \right) \epsilon^{\nu \tilde \Delta \Lambda \beta}\gamma_\beta   + \mu \leftrightarrow \nu\Biggl] \tau^c u(p_i,s_i),
\label{ParaM}
\end{eqnarray}
where any index $\rho \in \{ P, \tilde \Delta, \Lambda\}$ means contraction with the corresponding variable, e.g $ \epsilon^{\nu \tilde \Delta P \beta} =  \epsilon^{\nu \alpha \kappa \beta}\tilde \Delta_\alpha P_\kappa $.~The GTFFs $f_i$ are multiplied with separately conserved structures and they are functions of $\tilde t, t$ and $s$.~

  In general one can define $P$ and $\Lambda$ differently, such that the parametrization in Eq.~(\ref{ParaM}) remains the same, as long as they are orthogonal to $\tilde \Delta$, and $P$, $\tilde \Delta$ and $\Lambda$ are linearly independent.~In appendix \ref{ros} we show some formulas that we derived to reduce redundant structures that may appear while calculating the matrix element.
  
  \medskip
  
\begin{figure}[H]
	\centering
	\includegraphics[width=0.6 \textwidth]{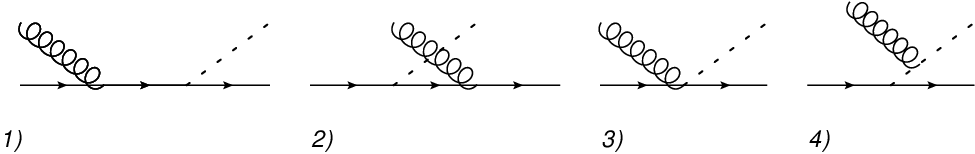}
	\caption{Tree-order diagrams contributing to the matrix element of OPGP.~Solid and dashed lines correspond to nucleons and pions, respectively, while curly lines represent gravitons.}
	\label{OPG:sp}
\end{figure}

           As an application we consider the tree-order contributions to the OPGP up to third order.~There are four tree-order diagrams shown in Fig.~\ref{OPG:sp}.
The corresponding contributions to the GTFFs are given in appendix \ref{ros-I}.~At one-loop order there appears a large number of diagrams, calculation and analysis of which are beyond the scope of this work.~Such calculation is under way and will be a subject of a separate publication.

\section{Summary and outlook} 
\label{III}
In this work we generalized the chiral effective Lagrangian in the presence of external gravitational field to include the delta resonances at the leading order and calculated the corresponding contributions to the nucleon GFFs.~From the full one-loop expressions we obtained the D-term and the corresponding slopes expanded in powers of small quantities.~To see the difference between the GFFs in the cases with and without $\Delta$ resonances we plotted the numerical results generated by the full expressions of the GFFs, which are too lengthy to be given explicitly in this work.~From the plots shown in Fig.~\ref{NGFF:sp} one can see, that the $\Delta$ resonances give significant contributions to the nucleon GFFs that increase with the momentum transfer.~Based on the numerical analysis at the end of section \ref{II} we argued that the contributions of the $\Delta$ resonances to the GFFs are much larger than the estimated error of our calculations.~This implies that the inclusion of the delta resonances as explicit degrees of freedom is essential.~This conclusion is also supported by the non-analytic parts of the GFFs in the chiral limit shown in Fig.~\ref{GFF-Non-Analytic}.

    From the non-analytic contributions to the nucleon GFFs we obtained the long-range behavior of the corresponding spatial densities in ZAMF of Ref.~\cite{Panteleeva:2023evj}.~We noticed, that the positivity of the energy density is broken by the terms that come from $\Delta$ coupling to nucleon and pion.~This suggests that a more detailed study between the stability conditions and decay of unstable particles is in order.

     In the second part we studied the matrix element of the EMT corresponding to gravitational $N \mapsto N \pi$ transition and gave a general parametrization in terms of twelve independent GTFFs.~This parametrization offers a simple and consistent way to do the calculations of the corresponding matrix element, where the current conservation and invariance under parity, charge and Hermitian conjugation is directly seen.~As an application we used the constructed Lagrangian to obtain the tree-order contributions to these GTFFs up to third chiral order.~One-loop order calculations within chiral EFT are underway.~Using arguments, that are similar to those given in section \ref{Sec:2}, one can find a parametrization for any non-diagonal matrix element of the EMT, e.g. when in the final state a photon is emitted instead of a pion or the final state contains one $\Delta$ and a pion or a photon.~
     
        The OPGP is related to deeply virtual hard exclusive processes and the corresponding GTFFs can be extracted from the data that will be available from future experiments.~From this connection we can obtain information on the unknown LECs, which will help testing the chiral EFT.

\acknowledgements

The author thanks E.~Epelbaum, J.~Gegelia, L.~Bovermann, J.~Panteleeva and L.~Ebener for helpful discussions and the comments on the draft. 
This work was supported in part by BMBF (Grant No. 05P21PCFP1), by
DFG and NSFC through funds provided to the Sino-German CRC 110
``Symmetries and the Emergence of Structure in QCD'' (NSFC Grant
No. 11621131001, DFG Project-ID 196253076 - TRR 110),
by ERC  NuclearTheory (grant No. 885150),  by the EU Horizon 2020 research and
innovation programme (STRONG-2020, grant agreement No. 824093),
and by the MKW NRW
under the funding code NW21-024-A.

\appendix

\section{Building blocks of the actions and expressions of EMT} 
\label{A-I}
The building blocks of the actions in this work are given as follows:
\begin{eqnarray}
 \overset{\leftrightarrow}{\nabla}_\mu 
 &= &
 \frac{1}{2}( \overset{\rightarrow}{\nabla}_\mu - \overset{\leftarrow}{\nabla}_\mu )
 \,,
 \nonumber\\
 \overset{\rightarrow}{\nabla}_\mu \Psi^i_\nu 
 & = &
 \nabla_\mu^{ij} \Psi^j_\nu = \left[
    \delta^{ij}\partial_\mu + \delta^{ij}\Gamma_\mu-i\delta^{ij} v_\mu^{(s)}-i \epsilon^{ijk}\text{Tr}\left(\tau^k\Gamma_\mu\right) 
    + \frac{i}{2}\delta^{ij}\omega^{ab}_\mu \sigma_{ab}\right]\Psi^j_\nu - \Gamma^{\alpha}_{\mu\nu}\Psi^i_\alpha, 
\nonumber \\ 
\bar\Psi^i_\nu \overset{\leftarrow}{\nabla}_\mu 
& = &
\nabla_\mu^{ij} \Psi^j_\nu =  \bar\Psi^j_\nu \left[
    \delta^{ij}\partial_\mu - \delta^{ij}\Gamma_\mu + i\delta^{ij} v_\mu^{(s)} + i \epsilon^{ijk}\text{Tr}\left(\tau^k\Gamma_\mu\right) 
    - \frac{i}{2}\delta^{ij}\omega^{ab}_\mu \sigma_{ab}\right] - \bar\Psi^i_\alpha\Gamma^{\alpha}_{\mu\nu}, 
\nonumber \\ 
\overset{\rightarrow}{\nabla}_\mu \Psi 
& = & 
\partial_\mu\Psi +\frac{i}{2} \, \omega^{ab}_\mu \sigma_{ab} \Psi + \left( \Gamma_\mu  -i v_\mu^{(s)}\right)\Psi\,,
 \nonumber\\
\bar\Psi \overset{\leftarrow}{\nabla}_\mu
& = &
\partial_\mu\bar\Psi -\frac{i}{2} \, \bar\Psi \, \sigma_{ab} \, \omega^{ab}_\mu - \bar\Psi \left( \Gamma_\mu  -i v_\mu^{(s)}\right) \,, 
\nonumber\\
\omega_\mu^{ab} 
&=&
-\frac{1}{2} \, g^{\nu\lambda} e^a_\lambda \left( \partial_\mu e_\nu^b
- e^b_\sigma \Gamma^\sigma_{\mu \nu} \right),
\nonumber\\
\chi_+ 
 & = & u^\dagger \chi u^\dagger+u \chi^\dagger u,
 \nonumber\\
\hat \chi_+ 
& = & \chi_+ -\frac{1}{2} \langle \chi_+\rangle,
\nonumber\\
\chi
&=& 2 B_0 (s+ip),
\nonumber\\
u_\mu & = & i \left[ u^\dagger \partial_\mu u  - u \partial_\mu u^\dagger -i (u^\dagger v_\mu u - u v_\mu u^\dagger )\right]\,,\nonumber\\
\Gamma_\mu & = & \frac{1}{2} \left[ u^\dagger \partial_\mu u  +u \partial_\mu u^\dagger -i (u^\dagger v_\mu u+u v_\mu u^\dagger )\right]~,\nonumber\\
u_{\mu,i}
&=& \frac{1}{2}  \text{Tr} \left(  u_{\mu} \tau^i  \right).
\label{Bb1}
\end{eqnarray}
 Using Eqs.~(\ref{EMTMatter}) and (\ref{EMTfermion}) we obtain the following expressions for the EMT corresponding to the actions given in section \ref{I} 
\begin{eqnarray}
T^{{(2)}}_{\pi , \mu\nu}  & = &  \frac {F_\pi^2}{4}\, {\rm Tr} ( D_\mu U  (D_\nu U)^\dagger)
-\frac{ \eta_{\mu\nu}}{2} \left\{ \frac {F_\pi^2}{4}\, {\rm Tr} ( D^\alpha U  (D_\alpha U)^\dagger )
+  \frac{F_\pi^2}{4}\,{\rm Tr}(\chi U^\dagger +U \chi^\dagger) \right\} + \left( \mu \leftrightarrow
\nu \right),
\label{PionEMT}
\end{eqnarray}
\begin{eqnarray}
T^{{  (1,2,3)}}_{\rm N  \pi, \mu\nu}  & = &  \frac{i}{2} \bar\Psi \,  \gamma_\mu  \overset{\leftrightarrow}{D}_\nu \Psi - \frac{\eta_{\mu\nu}}{2}\biggl(\, \bar\Psi \, i  \gamma^\alpha \overset{\leftrightarrow}{D}_\alpha \Psi -m \bar\Psi\Psi  \biggr) \nonumber
\\
&-&  \frac{c_2}{4 m^2} \left[  \langle u_\mu u^\beta\rangle  \left( \bar\Psi \left\{ D_\nu, D_\beta\right\}  \Psi+
 \left\{ D_\nu, D_\beta\right\}  \bar\Psi \Psi \right)
  \right] +  \frac{i c_2}{16 m^2} \, \partial^{{ \rho}} \left\{  \langle u^\alpha u^\beta\rangle  
   \left[
 D_\alpha  \bar\Psi \eta_{\beta\nu} \sigma_{{ \rho}\mu} \Psi  +
 D_\beta  \bar\Psi \eta_{\alpha\nu}  \sigma_{{ \rho}\mu} \Psi \right] \right. \nonumber\\
& - & \left.  2 \langle u^\alpha u^\beta\rangle 
  \bar\Psi \eta_{\beta\nu}  \sigma_{{ \rho}\mu}  D_\alpha  \Psi 
  \right\} + \frac{c_2}{4 m^2}  \left\{  \partial^\alpha \left[ \langle u_\alpha u_\mu \rangle  D_\nu ( \bar\Psi \Psi ) -\frac{1}{2} \langle u_\mu u_\nu \rangle  D_\alpha ( \bar\Psi \Psi )\right] \right\}
  \nonumber\\
  &+& \frac{1}{2} c_3 \, \bar\Psi \langle u_\mu u_\nu \rangle   \Psi 
 +\frac{i c_4}{8} \bar\Psi \left(  \sigma_{\nu\beta}\,\left[ u_\mu ,u^\beta\right] +\sigma_{\alpha\nu}\,\left[ u^\alpha ,u_\mu \right] \right)  \Psi +  \frac{c_6}{8m} \bar\Psi \sigma_{\nu\beta} F^+_{\mu\alpha} \eta^{\alpha\beta} \Psi 
\nonumber\\
&+&  \frac{c_7}{8m} \bar \Psi \sigma_{\nu\beta} \langle F^+_{\mu\alpha}\rangle \eta^{\alpha\beta} \Psi \nonumber
\\
&-&
  \frac{1}{2} \eta_{\mu\nu}  \left[ c_1 \langle \chi_+\rangle  \bar\Psi  \Psi  - \frac{c_2}{8 m^2} \langle u_\alpha u_\beta\rangle  \left( \bar\Psi \left\{ D^\alpha, D^\beta\right\}  \Psi  
+
 \left\{ D^\alpha, D^\beta\right\}  \bar\Psi \Psi \right) \right. \nonumber
 \\
 &+&
 \left. \frac{c_3}{2} \, \langle u_\alpha u^\alpha\rangle  \bar\Psi  \Psi
+ \frac{i c_4}{4} \, \bar\Psi\sigma^{\alpha\beta}\,\left[ u_\alpha ,u_\beta\right] \Psi + c_5 \bar\Psi\hat \chi_+ \Psi + \frac{c_6}{8m} \bar\Psi\sigma^{\alpha\beta} F^+_{\alpha\beta} \Psi + \frac{c_7}{8m} \bar \Psi \sigma^{\alpha\beta} \langle F^+_{\alpha\beta} \rangle \Psi 
 \right] 
 \nonumber
 \\
 &+&
 \left( \mu \leftrightarrow \nu \right)
     \,,
\label{MEMT}
\end{eqnarray}

\begin{eqnarray}
T^{{  (1)}}_{\Delta \pi  ,\mu\nu} & = &   -\Bar{\Psi}^{i}_\mu  \,  i  \gamma^\alpha \overset{\leftrightarrow}{D}_\alpha  \Psi^{i}_\nu + \Bar{\Psi}^{i}_\alpha  \,  i  \gamma^\alpha \overset{\leftrightarrow}{D}_\mu  \Psi^{i}_\nu   +
       \Bar{\Psi}^{i}_\mu  \,  i  \gamma^\alpha \overset{\leftrightarrow}{D}_\nu  \Psi^{i}_\alpha + m_\Delta \Bar{\Psi}^{i}_\mu   \Psi^{i}_\nu - \frac{i}{2} \, \bar\Psi^i_\alpha \,  \gamma_\mu  \overset{\leftrightarrow}{D}_\nu \Psi^{i\alpha}
\nonumber\\
& + & \, \frac{i}{2}   \biggl(  \bar\Psi^i_\mu \,  \gamma_\nu  \overset{\leftrightarrow}{D}_\alpha \Psi^{i\alpha} + \bar\Psi^{i\alpha} \,  \gamma_\nu \overset{\leftrightarrow}{D}_\alpha \Psi^{i}_\mu   -  \bar\Psi^i_\mu \,  \gamma_\nu \gamma^\alpha\gamma_\beta  \overset{\leftrightarrow}{D}_\alpha \Psi^{i,\beta}  - \bar\Psi^i_\alpha \gamma^\alpha \gamma_\nu \gamma^\beta  \overset{\leftrightarrow}{D}_\mu \Psi^{i}_\beta
- \bar\Psi^i_\alpha \gamma^\alpha\gamma^\beta \gamma_\nu  \overset{\leftrightarrow}{D}_\beta \Psi^{i}_\mu  \biggl)
\nonumber\\
&+& \frac{i}{4}\,\partial^\lambda \biggl[\bar\Psi^{i, \alpha}  \biggl( \gamma_\mu \eta_{\lambda [\alpha}\eta_{\beta] \mu} +\eta_{\lambda \mu}\eta_{\nu [\alpha}\gamma_{\beta]} + \eta_{\mu\nu} \eta_{\lambda[\beta}\gamma_{\alpha]} \biggr)
\Psi^{i,\beta}   
\biggr] - \frac{m_\Delta}{2} \, \left( \bar\Psi^i_\mu \,  \gamma_\nu \gamma^\alpha \Psi^{i}_\alpha + \bar\Psi^i_\alpha \,  \gamma^\alpha \gamma_\nu \Psi^{i}_\mu \right)  
\nonumber\\
&+& \frac{\eta_{\mu\nu} }{2} \biggl[  
     \, \Bar{\Psi}^{i}_\alpha  \,  i \gamma^\beta \overset{\leftrightarrow}{D}_\beta  \Psi^{i\alpha } -  m_\Delta \, \Bar{\Psi}^{i}_\alpha   \Psi^{i\alpha}  -\Bar{\Psi}^{i}_\alpha  i \gamma^{\alpha}{\overset{\leftrightarrow}{D}_\beta}  \Psi^{i\beta} -\Bar{\Psi}^{i\alpha}i \gamma^{\beta}{\overset{\leftrightarrow}{D}_\alpha}  \Psi^{i}_\beta 
 + i  \Bar{\Psi}^{i}_\rho \gamma^\rho \gamma^\alpha\gamma^\lambda \overset{\leftrightarrow}{D}_\alpha \Psi^{i}_\lambda  \nonumber\\ &+&  m_\Delta \Bar{\Psi}^{i}_\alpha \gamma^\alpha \gamma^\beta  \Psi^{i}_\beta \biggr] 
   +  \left( \mu \leftrightarrow \nu \right)  \,,
   \label{EMTB}
\\
T^{(1)}_{\pi N\Delta, \mu\nu}  &=&g_{\pi N\Delta} \biggl\{  \frac{1}{2} ~\eta_{\mu \nu} \left[ \bar \Psi^i_\alpha u^{\alpha}_i\Psi + \bar \Psi u^{\alpha}_i   \Psi^i_\alpha - \bar \Psi^i_\alpha \gamma^\alpha \gamma^\beta u_{\beta }^i \Psi - \bar \Psi  \gamma^\beta  \gamma^\alpha u_{\beta }^i  \Psi^i_\alpha \right] -  \bar\Psi^i_\mu u_{\nu}^i\Psi -  \bar\Psi u_{\nu}^i \Psi^i_\mu  \nonumber
\\
&+&\frac{1}{2}  ~ \left[ \bar \Psi^i_\mu \gamma_\nu  \gamma^\alpha u_{\alpha }^i \Psi + \bar \Psi^i_\alpha \gamma^\alpha  \gamma_\mu u_{\nu}^i \Psi + \bar \Psi  \gamma^\alpha \gamma_\nu  u_{\alpha }^i  \Psi^i_\mu + \bar \Psi  \gamma_\mu  \gamma^\alpha u_{\nu}^i  \Psi^i_\alpha \right] \biggl\} +  \left( \mu \leftrightarrow \nu \right)  \, ,
\\
T^{2, \mu\nu}_{\rm \pi N} &=& \frac{c_8}{8} \left[\eta_{\mu \nu} \partial^2 - \partial_\mu \partial_\nu \right]  \bar\Psi \Psi  + \frac{ i c_9}{m} \left[ \partial^2  \left( \bar\Psi \gamma_\mu
 \overset{\leftrightarrow}{\nabla}_\nu \Psi \right)+ \eta_{\mu \nu} \partial^\alpha \partial^\beta  \left(  \bar\Psi \gamma_\alpha
\overset{\leftrightarrow}{\nabla}_\beta \Psi \right) \right.
\nonumber
\\
&-& 
\left. \partial^\alpha \partial_\mu \left(  \bar\Psi \gamma_\alpha
\overset{\leftrightarrow}{\nabla}_\nu \Psi + \bar\Psi \gamma_\nu
\overset{\leftrightarrow}{\nabla}_\alpha \Psi \right)\right] 
+  \left( \mu \leftrightarrow \nu \right)  \,,
\\
T^{4, \mu\nu}_{\rm \pi} &=& \left(\eta_{\mu \nu} \nabla^2 - \nabla_\mu \nabla_\nu \right) \left[ l_{11} \, {\rm Tr} ( D^\alpha U  (D_\alpha U)^\dagger ) +  l_{13} \,{\rm Tr}(\chi U^\dagger +U \chi^\dagger)  \right] + \frac{l_{12}}{2} \left[ \partial^2  \left( {\rm Tr} ( D_\mu U  (D_\nu U)^\dagger ) \right)\right.
\nonumber
\\
&+&
\left. 
\eta_{\mu \nu} \partial^\alpha \partial^\beta  \left( {\rm Tr} ( D_\alpha U  (D_\beta U)^\dagger )\right) -\partial^\alpha \partial_\mu \left( {\rm Tr} ( D_\alpha U  (D_\nu U)^\dagger )+
 {\rm Tr} ( D_\nu U  (D_\alpha U)^\dagger )\right)\right]
+  \left( \mu \leftrightarrow \nu \right)  \,,
\\
T^{3, \mu\nu}_{\rm \pi N} &=& -\frac{1}{4} \eta_{\mu\nu} \left[\tilde d_{10} \bar \Psi  \gamma^\alpha \gamma_5  \langle u^2 \rangle u_\alpha \Psi + \tilde d_{16} \bar \Psi  \gamma^\alpha \gamma_5  \langle \chi_+ \rangle u_\alpha \Psi  +\tilde d_{17} \bar \Psi  \gamma^\alpha \gamma_5  \langle \chi_+ u_\alpha \rangle  \Psi + i ~ \tilde d_{18} \bar \Psi  \gamma^\alpha \gamma_5  \left[ D_\alpha ,  \chi_- \right]  \Psi   \right]
\nonumber
\\
&+& 
\frac{1}{4} \left[ \tilde d_{10} \bar \Psi  \gamma_\mu \gamma_5  \langle u^2 \rangle u_\nu \Psi + \tilde d_{16} \bar \Psi  \gamma_\mu \gamma_5  \langle \chi_+ \rangle u_\nu \Psi  +\tilde d_{17} \bar \Psi  \gamma_\mu \gamma_5  \langle \chi_+ u_\nu \rangle  \Psi  +  i ~ \tilde d_{18} \bar \Psi  \gamma_\mu \gamma_5  \left[ D_\nu ,  \chi_- \right]  \Psi  \right]
\nonumber
\\
&+&\tilde d_{g1} \left[\eta_{\mu \nu} \partial^2 - \partial_\mu \partial_\nu \right]  \bar\Psi \gamma^\alpha
\gamma_5 u_\alpha \Psi  
\nonumber
\\
&+&
\frac{\tilde d_{g2}}{2 } \left[ \partial^2  \left( \bar\Psi \gamma_\mu
\gamma_5 u_\nu \Psi \right)+ \eta_{\mu \nu} \partial^\alpha\partial^\beta  \left(  \bar\Psi \gamma_\alpha
\gamma_5 u_\beta \Psi \right) -
 \partial^\alpha\partial_\mu \left(  \bar\Psi \gamma_\alpha
\gamma_5 u_\nu \Psi + \bar\Psi \gamma_\nu
\gamma_5 u_\alpha \Psi \right)\right] 
\nonumber
\\
&+& \frac{1}{2} \partial
^{\kappa}  \partial^{\beta} \left(   {\mathcal{A}_{\beta \mu \kappa\nu}}-{\mathcal{A}_{\beta \kappa \mu\nu}} -{\mathcal{A}_{\mu \nu \kappa\beta}}+{\mathcal{A}_{\mu \kappa \nu\beta}}\right)  - \tilde d_{g5} \left[\eta_{\mu \nu} \partial^2\partial^\alpha  - \partial_\mu \partial_\nu \partial^\alpha  \right] \left( \bar\Psi \overset{\leftrightarrow}{\nabla}_\alpha  \Psi  \right)
\nonumber
\\
&+&   \left( \mu \leftrightarrow \nu \right)  \,,
\end{eqnarray}
where
\begin{eqnarray} 
\mathcal{A}_{\mu \lambda \nu \rho} \equiv  \tilde d_{g3}~ \bar \Psi  \sigma_{\rho \mu} \gamma_5 u_\lambda \overset{\leftrightarrow}{\nabla}_\nu \Psi -  i  \tilde d_{g4}~ \partial_\nu \left(  \bar \Psi \sigma_{\rho \mu} \overset{\leftrightarrow}{\nabla}_\lambda \Psi \right)\, ,
\end{eqnarray}
 $\eta_{\mu\nu}$ is the Minkowski metric tensor with the
signature $(+,-,-,-)$ and $A^{[\alpha} B^{\beta]} = A^{\alpha} B^{\beta} -  A^{\beta}B^{\alpha}$,  $A^{(\alpha} B^{\beta)}
= A^{\alpha} B^{\beta} +  A^{\beta}B^{\alpha}$.~The covariant derivatives ${D}$ acting on spin-1/2 and spin-3/2 fields coincide with $\nabla$ in Eq.~(\ref{Bb1}) with $g_{\mu \nu} = \eta_{\mu\nu}$, i.e. $ \Gamma_{\mu\nu}^\beta = \omega_{\mu}^{ab} = 0$.

\section{Variation of terms in the action with gravity}
\label{A-II}
To obtain the EMT we have to deal with variations of various quantities, some of them are of the following form
\begin{eqnarray}
\delta \mathcal{L}_R=\int d^4 x \sqrt{- g} {A^{\mu \lambda \nu}}_{\rho} \, \delta  {R^{\rho}}_{\mu \lambda \nu} |_{g = \eta}\, ,
\end{eqnarray}
where ${A^{\mu \lambda \nu}}_{\rho}$ is some tensor that depends on the fields and/or their covariant derivatives.~From Eq.~($4.60$) in Ref.~\cite{Carroll:1997ar} we have: 
\begin{align}
\delta  {R^{\rho}}_{\mu \lambda \nu} = \nabla_{\lambda} \delta \Gamma^{\rho}_{\mu \nu}-\nabla_{\nu} \delta \Gamma^{\rho}_{\lambda \mu} = \left( \delta^{\kappa}_{\lambda}\delta^{\omega}_{\nu}- \delta^{\kappa}_{\nu}\delta^{\omega}_{\lambda}\right)\nabla_{\kappa} \delta \Gamma^{\rho}_{\mu \omega},
\end{align}
i.e. 
\begin{eqnarray}
\delta \mathcal{L}_R &=& \int d^4 x \sqrt{- g} {A^{\mu \lambda \nu}}_{\rho}  \left( \delta^{\kappa}_{\lambda}\delta^{\omega}_{\nu}- \delta^{\kappa}_{\nu}\delta^{\omega}_{\lambda}\right)\nabla_{\kappa} \delta \Gamma^{\rho}_{\mu \omega} |_{g = \eta}
\nonumber
\\
&=&\int d^4 x \sqrt{- g}  \delta \Gamma^{\rho}_{\mu \omega} \nabla_{\kappa}  \left( {A^{\mu \omega \kappa}}_{\rho}-{A^{\mu \kappa \omega}}_{\rho} \right)|_{g = \eta}\,.
\end{eqnarray}
From Eq.~($4.59$) in the same reference we derive  the following expression for the variation of the Christoffel symbols 
\begin{align}
\delta \Gamma^{\rho}_{\mu \omega} = -\frac{1}{2} \left[ g_{\lambda \omega} \nabla_{\mu} \delta g ^{\lambda \rho} + g_{\lambda \mu} \nabla_{\omega} \delta g ^{\lambda \rho}-  g_{\mu \alpha} g_{\omega \beta} \nabla^{\rho} \delta g ^{\alpha \beta} \right],
\end{align}
i.e. 
\begin{eqnarray}
\delta \mathcal{L}_R &=& -\frac{1}{2}  \int d^4 x \sqrt{- g}  \left[ g_{\lambda \omega} \nabla_{\mu} \delta g ^{\lambda \rho} +  g_{\lambda \mu} \nabla_{\omega} \delta g ^{\lambda \rho}-  g_{\mu \alpha} g_{\omega \beta} \nabla^{\rho} \delta g ^{\alpha \beta} \right]\nabla_{\kappa}  \left( {A^{\mu \omega \kappa}}_{\rho}-{A^{\mu \kappa \omega}}_{\rho} \right)|_{g = \eta}
\nonumber
\\
&=& \frac{1}{2}  \int d^4 x \sqrt{- g}  \left( \nabla
^{\kappa}  \nabla^{\beta} \left[   {A_{\beta \lambda \kappa\rho}}-{A_{\beta \kappa \lambda\rho}}-{A_{\lambda \rho \kappa\beta}}+{A_{\lambda \kappa \rho\beta}}\right] \right) \delta g ^{\lambda \rho}|_{g = \eta}.\label{b6}  \end{eqnarray}
 We can use the last equation to derive the EMT contributions coming from the Ricci tensor as follows: 
 \begin{eqnarray}
\int d^4 x \sqrt{- g} B^{\mu\nu} \delta R_{\mu\nu} |_{g = \eta}  = \int d^4 x \sqrt{- g} B^{\mu\nu} \delta ( g_{\rho}^{\lambda}  {R^{\rho}}_{\mu \lambda \nu} ) |_{g = \eta}  = \int d^4 x \sqrt{- g} B^{\mu\nu} g_{\rho}^{\lambda} \delta {R^{\rho}}_{\mu \lambda \nu}|_{g = \eta}  \,.
\label{b7}
 \end{eqnarray}
By taking $ {A^{\mu \lambda \nu}}_{\rho} = B^{\mu\nu} g_{\rho}^{\lambda}$ and using Eq. (\ref{b6}) we obtain:
  \begin{eqnarray}
\int d^4 x \sqrt{- g} B_{\mu\nu} \delta R^{\mu\nu}|_{g = \eta} &=& 
\frac{1}{2} \int d^4 x \sqrt{- g}  \left[ \nabla^2 B_{\lambda \rho} + g_{\lambda \rho } \nabla^{\kappa} \nabla^{\beta} B_{\beta \kappa} - \nabla^{\beta}\nabla_{\rho} (  B_{\beta \lambda}+ B_{\lambda\beta }  ) \right] \delta g ^{\lambda \rho}|_{g = \eta}\,. \label{8} 
 \end{eqnarray}
We can use the last equation to derive contributions coming from Ricci scalar as follows: 
 \begin{eqnarray}
\int d^4 x \sqrt{- g} V \delta R |_{g = \eta} &=& \int d^4 x \sqrt{- g} V g_{\alpha \beta} \delta R^{\alpha \beta} |_{g = \eta} = \int d^4 x \sqrt{- g}  \left[ g_{\lambda \rho} \nabla^2 V  - \nabla_{\lambda}\nabla_{\rho} V  \right] \delta g ^{\lambda \rho}|_{g = \eta} \,,
\label{9}
  \end{eqnarray}
where $ B^{\mu\nu}$ and $V$ are arbitrary tensor and scalar fields, respectively.
\section{Reduction of structures}
\label{ros}
We make use of various equations from appendix B of Ref.~ \cite{Cotogno:2019vjb} to derive useful relations to get rid of some redundant terms in the amplitude of the OPGP.~The relations we obtained are given by
\begin{eqnarray}
\gamma^\mu \gamma_5 &\overset{.}{=} &\frac{m_N}{t~\tilde t } \left( \tilde \Delta^\mu \left[ \tilde t + \delta t \right] -\delta t~ \lambda^\mu + \tilde t P^\mu \right) \gamma_5 \nonumber
\\
&+&i \frac{\delta s -\tilde t}{c} \Biggl\{ \left(\delta s - \tilde t \right)\left[\tilde t P^\mu - \delta t ~\Lambda^\mu  + (\tilde t + \delta t) \tilde \Delta^\mu \right] \epsilon^{\tilde \Delta P \Lambda \beta}
+2 \tilde t t \left[ \tilde t \epsilon^{\nu P \Lambda \beta} \right. \nonumber
\\
&+&
\left. \left(\delta t +2 \delta s - \tilde t \right)  \epsilon^{\nu \tilde \Delta \Lambda \beta}  \right]
\Biggl\} \gamma_\beta, \,
\\
\tilde{ \slashed \Delta} \gamma_5 &\overset{.}{=} & \frac{1}{2 \tilde t t} \Biggl\{  2 \tilde t m_N (\tilde t +\delta t)\gamma_5 + i (\delta s - \tilde t) \epsilon^{\tilde \Delta P \lambda \beta}\gamma_\beta \Biggl\},\,
\\
\tilde{ \slashed \Delta} \gamma^\mu \gamma_5 &\overset{.}{=} & \tilde \Delta^\mu \gamma_5 + \frac{\delta s - \tilde t}{2 t} \left( P^\mu + \Lambda^\mu \right) \gamma_5 + i  \frac{\tilde t (\delta s - \tilde t)^2 }{2 m_N ~c} \left[ 2 t \left(\Lambda^\mu -\tilde \Delta^\mu  \right) + (\delta s -\tilde t) \left( P^\mu + \Lambda^\mu \right)\right] \epsilon^{\tilde \Delta P \lambda \beta} \gamma_\beta 
\nonumber
\\ 
&+&  i  \frac{\tilde t t (\delta s - \tilde t) }{ m_N ~c} \left[\tilde t (\delta s - \tilde t )\epsilon^{\mu P \Lambda \beta } + 2 \left((\delta s - \tilde t)(\delta s + \delta t) + \tilde t t \right) \epsilon^{\mu \tilde \Delta \Lambda \beta }  \right]\gamma_\beta \, , 
\\
\epsilon^{\mu \tilde \Delta P \beta }\gamma_\beta &=&\frac{2 \tilde t t (\delta s -{ \tilde \Delta}^2 )}{c } \left( \tilde \Delta^\mu \left[ \tilde t + \delta t \right] -\delta t~ \lambda^\mu + \tilde t P^\mu \right) \epsilon^{\tilde \Delta P \Lambda \beta }\gamma_\beta \nonumber
\\
&+& \frac{2 \tilde t^2 t \left( 2 t (-2 \tilde t +2 \delta s + \delta t) - (\delta s -\tilde t)(\tilde t + \delta t) \right) }{c}\epsilon^{\mu \tilde \Delta \Lambda \beta }\gamma_\beta - \frac{2 \tilde t^2 t (\delta s -\tilde t) (\tilde t + \delta t )}{c} \epsilon^{\mu P \Lambda \beta }\gamma_\beta\,,
\end{eqnarray}
where $c = 2 \tilde t^2 t \left[ 2 \tilde t t + (\delta s -\tilde t)(\delta t +\tilde t) \right]$ and the symbol $ \overset{.}{=} $ means both sides are sandwiched by spinors and the on-shell relations are used.~For example, the Gordon identity
\begin{eqnarray}
\bar u(p_f,s_f) \gamma^\mu u(p_i,s_i) = \frac{1}{2 m} \bar u(p_f,s_f)\left[ p_f^\mu+p_i^\mu +i \sigma^{\mu \nu} (p_{f \nu}-p_{i \nu}) \right] u(p_i,s_i), 
\end{eqnarray}
can be written as
\begin{eqnarray}
\gamma^\mu\overset{.}{=}  \frac{1}{2 m} \left[ p_f^\mu+p_i^\mu +i \sigma^{\mu \nu} (p_{f \nu}-p_{i \nu}) \right]. 
\end{eqnarray}

\section{Tree-order expressions to GTFFs}
\label{ros-I}
The diagrams in Fig.~\ref{OPG:sp} yield the following contributions to the GTFFs
\begin{eqnarray}
f_1 &=& -\frac{1}{2 F t \delta t ~\delta u}\biggl\{ 2 g_A  c_9 (M_\pi^4-t^2)+  \tilde d_{g2}~ m_N \delta t \delta u -4 l_{13} g_A \frac{m_N t \delta u}{F_\pi^2}  +  \tilde d_{g3} \frac{\delta t \delta u \delta s (\delta t + \delta u) }{2 \tilde t}\nonumber
\\
&-& \tilde d_{g4} g_A \delta t (\delta u (\delta t +\delta u) - 4 m_N^2 \tilde t)  +\left(  \left[ 2 d_{16} - d_{18} \right]  M_\pi^2  + g_A \right) m_N \frac{(2t -\delta t ) (\delta t + \delta u)}{\tilde t } \biggl\} \, ,
\\
f_2 &=&  \frac{1}{ F t \delta u \tilde t \left[ (\tilde t + \delta t) (\delta t + \delta u) - 2 \tilde t t \right]}\biggl\{ \frac{1}{4}   \left[2 g_A  c_9 \delta t + \tilde d_{g2}~ m_N \delta u  \right]     (\delta t +\delta u)^2 \tilde t \nonumber
\\
&+& \frac{1}{8} \tilde d_{g3} \delta u \delta s (\delta t + \delta u)^3 +\frac{1}{4}\tilde d_{g4} g_A \tilde t \left(- (\delta t + \delta u)^3 \delta u + 4 m_N^2 \left[ - \tilde t ~t^2 + (M_\pi^2 - \delta u )^2 \tilde t \right]  \right) 
\nonumber
\\
&-&  \frac{1}{4} \left(  \left[ 2 d_{16} - d_{18} \right] M_\pi^2 + g_A \right)  m_N (\delta t + \delta u)^3
\biggl\} \, ,
\\
f_3 &=& -\frac{1}{4 F t \delta s \tilde t^2} \biggl\{ 4 g_A c_9 \left(\tilde t (t^2-M_\pi^4)  + (\delta t + \delta u) (\tilde t ^2 + M_\pi^4 - t^2 -2 \tilde t ~t) \right)+ 2 m_N \tilde d_{g2}~\delta s \delta t^2
\nonumber
\\
&-&
 \tilde d_{g3} \delta s \delta u  \tilde t (\delta s - \tilde t) -  2\tilde d_{g4} g_A \left(\delta s (\delta t +2 \delta s- 2 \tilde t) (\delta s - \tilde t) + 4 m_N^2 \delta t (2 \delta s - \tilde t) \right)\tilde t
  \nonumber
\\
&-& 8  \frac{g_A m_N l_{13}}{F_\pi^2}t \delta t \delta s + 2 \left(  \left[ 2 d_{16} - d_{18} \right]  M_\pi^2 + g_A \right) m_N (2t - \tilde t) (\tilde t- \delta s)
  \biggl\}\, ,
 \end{eqnarray}
 \begin{eqnarray}
f_4 &=&  \frac{1}{ 8 F t \tilde t^2 \delta s ((\delta s  - \tilde t) (\tilde t + \delta t) + 2 \tilde t t) } \biggl\{- 4 g_A c_9  (\tilde t - \delta s)^2 \left(\tilde t^3 - \delta s \delta t^2 - \tilde t^2 (\delta s + 2 t) \right)
\nonumber
\\
&+&\left[  \tilde d_{g3} \tilde t  \delta u (\delta s - \tilde t +2 t) -2 \tilde d_{g2} m_N \delta t^2\right]  \delta s (\tilde t - \delta s)^2 
\nonumber
\\
&-&
2 \tilde t g_A \tilde d_{g4}  (\delta s - \tilde t) \left( -4 \delta t m_N^2 (\tilde t^2 - 3 \tilde t \delta s +2 \delta s (\delta s + t) ) - \delta s (\delta s - \tilde t) (-2 \tilde t + 2 \delta s + \delta t ) (-\tilde t + \delta s + 2t) \right)
\nonumber
\\
&-& 2 \tilde t \left( \left[ 2 d_{16} - d_{18} \right]  M_\pi^2 + g_A \right)  m_N (\delta s - \tilde t)^2( \delta s -\tilde t + 2t)
\biggl\} \, ,
\\
f_5 &=& -\frac{1}{8F t \delta s \delta u \tilde t} \biggl\{- 4 g_A c_9 \left((\delta s - \tilde t) (\tilde t^2 + 2 \delta s \delta t - \tilde t \delta t) + 2 \delta s M_\pi^4 - 2 \delta s t^2 - 4 \delta s t (\delta s- \tilde t) \right)
\nonumber
\\
&-& 4 m_N \tilde d_{g2}  \delta s \delta t (\tilde t - \delta s - \delta t) + \tilde d_{g3} \delta s (\tilde t - \delta s) (\tilde t - \delta t) (\tilde t - \delta s - \delta t)
\nonumber
\\
&-&16 \frac{g_A m_N l_{13}}{F_\pi^2} \delta s t (\delta t + \delta s - \tilde t) + 2  \left(  \left[ 2 d_{16} - d_{18} \right]  M_\pi^2 + g_A \right) m_N (\delta s - \tilde t) (\tilde t- \delta t)
\, ,
\nonumber
\\
&+&2g_A \tilde d_{g4} \left[ \tilde t ^2 ( -4 m_N^2 (\delta_1-2\delta u) - \delta u \delta_1 ) + \tilde t \delta_1 (4 m_N^2 (\delta_1 -3 \delta u) -\delta u^2) +\delta u \delta_1^2 (\delta u + \delta_1) \right]
\biggl\}\, ,
\end{eqnarray}
\begin{eqnarray}
f_6 &=& -  \frac{1}{\delta u F \tilde t  t \delta s ( (\tilde t + \delta t) (\delta t + \delta u) -  2 \tilde t t )} \biggl\{ \frac{g_A c_9}{4} (\tilde t - \delta s)^2 \left[- \tilde t^3 + (\tilde t^2 + 2 \delta s \delta t ) (\delta s + \delta t) 
\right.
\nonumber
\\
&+&
\left.  \tilde t \delta s (2t - 3 \delta t)\right] - \tilde d_{g2}  \delta s \delta t m_N \frac{(\delta s - \tilde t)^2}{4} (\delta t + \delta s - \tilde t)
\nonumber
\\
&-&\tilde d_{g3} \frac{\delta s (\delta s - \tilde t)^2}{16} (\delta s + \delta t - \tilde t) \left[ (\delta s- \tilde t) (\tilde t-\delta t )+ 2 \delta s t \right] \nonumber
\\
&+&
 \frac{\left(  \left[ 2 d_{16} - d_{18} \right]  M_\pi^2 + g_A \right) m_N }{8 } (\delta s - \tilde t)^2\left[ (\delta s- \tilde t) (\delta t -\tilde t) - 2 \delta s t \right]
\nonumber
\\
&+&\frac{g_A \tilde d_{g4}}{8}  \delta_1 \left[\tilde t ^2 ( \delta_1-2t) (4 m_N^2 (  \delta_1-2 \delta u) + \delta u  \delta_1)+ \tilde t \delta_1 (\delta u  \delta_1 (\delta u + 2 t) \right.\nonumber
\\
&-&
\left. 4 m_N^2 (4 \delta u t -2 t  \delta_1 + \delta_1^2 -3 \delta u  \delta_1) ) - \delta u  \delta_1^3 (\delta u + \delta_1)\right]
\biggl\}\, ,
\\
f_7 &=& \frac{1}{2 F \delta t \tilde t^2}  \biggl\{\frac{1}{t \delta u  \delta s} \left[ g_A  \left( \delta u (\delta_2 + \tilde t) ( \delta_2+ \delta u) \left[ -2 \delta_2 c_9 (\delta_2 + 2 \tilde t)- t \tilde t (c_8 +2 \tilde d_{g5} (\delta_2+2 \delta u)) +4 \delta_2 c_9 t\right] \right. \right.
\nonumber
\\
&+&
\left. \left. 
\delta_2 c_8 \tilde t m_N^2 (\delta_2+ \tilde t) (\delta_2 + 2 \tilde t -2 t) \right)
-2 \left(  \left[ 2 d_{16} - d_{18} \right]  M_\pi^2 + g_A \right) m_N  \delta u \delta_2 t (\delta_2 + \delta u)\right]
\nonumber
\\
&-&\frac{4 g_A m_N }{F_\pi^2} \left[ \tilde t^2 (4 l_{11} + l_{13}) + \tilde t (4 l_{11} \delta t + 8 l_{12} M_\pi^2  - 8 l_{11} t) - l_{13} \delta t^2 \right] - 8 m_N \tilde d_{g1} \delta t \tilde t
\nonumber
\\
&-&
m_N\frac{\delta t (\delta t + \tilde t)}{t} \left[\tilde d_{g2} \delta t - \tilde t (4 \tilde d_{g1} +\tilde d_{g2} ) \right] \biggl\}\, ,
\\
f_8 &=& \frac{\tilde t-\delta s }{4 \tilde t^2 \delta s F t \delta u} \biggl\{ g_A (\tilde t - \delta t ) \left( c_8 \tilde t m_N^2 -2 c_9 \delta s (\delta s +\delta t - \tilde t) \right) 
-
 \delta s m_N (4\tilde d_{g1} \tilde t +\tilde d_{g2}(\tilde t - \delta t) ) \delta u
\biggl\} \, ,
\\
f_9 &=& \frac{\tilde t - \delta s }{8 \tilde t F m_N \delta u ( \tilde t ( \tilde t +  \delta t - 2 t) - \delta s ( \tilde t + \delta t  )   ) } \biggl \{ 
2 \left[ 2 d_{16} - d_{18} \right] m_N M_\pi^2 (\tilde t- \delta s) 
\nonumber
\\
&+& 2 g_A \left( \tilde d_{g4} \tilde t ( 4 \delta t m_N^2 + (\delta s - \tilde t) (\delta t + \delta s - \tilde t) ) -2 c_9 \tilde t \delta t + m_N (\tilde t - \delta s) \right)
-
\delta u (  \tilde d_{g3} \delta s (\tilde t - \delta s ) + 2  \tilde d_{g2} \tilde t m_N )
\biggl\} \, ,
\\
f_{10 }&=&
-\frac{(\delta s- \tilde t)}{8 \tilde t \delta s F m_N \left(  (\delta s-\tilde t) (\tilde t +\delta t) + 2 \tilde t  t \right) }
\biggl \{ 
\left( 2 \left[ 2 d_{16} - d_{18} \right] m_N M_\pi^2  +   \tilde d_{g3} \delta s (\delta s + \delta t - \tilde t) \right)  (\delta s - \tilde t )
\nonumber
\\
&+& 2 g_A \left( m_N - \tilde d_{g4} \delta s(\delta t-2\tilde t + 2\delta s)  \right) (\delta s  - \tilde t) + 4 g_A (c_9 - 2  \tilde d_{g4} m_N^2) (\tilde t^2 + \delta s (\delta t- \tilde t)) -2  \tilde d_{g2} \delta s \delta t m_N
\biggl\} \, ,
\end{eqnarray}
\begin{eqnarray}
f_{11} &=&  
\frac{\delta s - \tilde t}{4 F m_N \delta s \tilde t (2 t \tilde t +(\delta s - \tilde t) (\delta t + \tilde t) )} \biggl\{
2 g_A c_9 (\tilde t \delta s (2 \delta s + 3 \delta t) -\delta s \delta t (2 \delta s + \delta t) - 2 \tilde t^2 (\delta s - M_\pi^2) )
\nonumber
\\
&+&
 \tilde d_{g2} m_N \delta s \delta t ( \delta t + 2 \delta s - \tilde t) -  \tilde d_{g3} \delta s (\delta s + \delta t - \tilde t) (t \tilde t + (\delta s- \tilde t) (\delta s + \delta t) )
\nonumber
\\
&+&
2 g_A \tilde d_{g4} \left( \delta s (2 \delta s + \delta t) (\delta s (\delta s + \delta t) + 2 \delta t m_N^2) +2 \tilde t^2 (\delta s + 2 m_N^2) (\delta s - M_\pi^2) \right.
\nonumber
\\
&-&
\left. \tilde t \delta s ((\delta s + \delta t) (4 \delta s+ \delta t) +m_N^2 (4 \delta s+ 6\delta t) - t (2\delta s + \delta t)) \right) 
\nonumber
\\
&+&
 2 \left(  \left[ 2 d_{16} - d_{18} \right]  M_\pi^2 + g_A \right) m_N (\tilde t (\delta s - M_\pi^2) -  \delta s(\delta s + \delta t))  
\biggl\}
 \, ,
\\
f_{12} &=&\frac{1}{4 F m_N \delta u \tilde t (-2 t \tilde t + (\delta t + \delta u) (\delta t + \tilde t) )} \biggl\{
2 g_A c_9 \tilde t (\delta t+ \delta u) ( \delta t \delta u  - \tilde t (\delta u - 2 M_\pi^2))
\nonumber
\\
&+&
 \tilde d_{g2} m_N \delta u \tilde t (\delta t + \delta u) (\delta t + 2 \delta u - \tilde t) +  \tilde d_{g3} \delta s (\delta s - \tilde t ) (\delta s + \delta t - \tilde t) ( (\delta s- \tilde t) (\delta s + \delta t) + t \tilde t  )
\nonumber
\\
&-&
2 g_A \tilde d_{g4} \tilde t (\delta t + \delta u ) \left[ 2 m_N^2 (\delta t \delta u - \tilde t ( \delta u - 2 M_\pi^2 ) ) + \delta u ( (\delta t + \delta u) (\delta u - \tilde t) +t \tilde t ) \right] \nonumber
\\
&+&
 2 \left(  \left[ 2 d_{16} - d_{18} \right]  M_\pi^2 + g_A \right) m_N  (\delta s- \tilde t) ( (\delta s- \tilde t) (\delta s + \delta t ) + t \tilde t )
\biggl\}   \, ,
\end{eqnarray}
where
\begin{eqnarray}
&& \delta s = s - m_N^2,\,  \quad \delta t = t - M_\pi^2,\, \quad \delta u =  \tilde t -\delta s - \delta t\,,
\nonumber
\\
&& \delta_1 = \delta t + \delta u,\,  \quad \delta_2 = \delta t - \tilde t.
\end{eqnarray}

\end{document}